\documentclass[usenatbib]{mnras}
\usepackage{verbatim,bm}
\usepackage[T1]{fontenc}
\usepackage{ae,aecompl}
\usepackage{graphicx}
\usepackage{amsmath,amssymb}
\usepackage{color,ulem}

\definecolor{webgreen}{rgb}{0,.5,0}
\definecolor{webbrown}{rgb}{.6,0,0}

\defcitealias{pra15}{PSB15}
\hypersetup{colorlinks=true,
            breaklinks=true,
            plainpages=false,
            bookmarksnumbered,
            bookmarksopen=true,
            bookmarksopenlevel=1,
            urlcolor=webbrown,
            linkcolor=blue,
            citecolor=webgreen
           }
\begin{document}
\title[Stochastic cold mode feedback with AGN jets]{AGN jet-driven stochastic cold accretion in cluster cores}
\author[Prasad, Sharma, \& Babul]{Deovrat Prasad$^{1}$\thanks{{deovrat@physics.iisc.ernet.in}},
Prateek Sharma$^{1}$\thanks{{prateek@physics.iisc.ernet.in}}
and Arif Babul$^{2,3,4}$\thanks{{babul@uvic.ca}}\\
$^{1}$Department of Physics \& Joint Astronomy Programme,Indian Institute of Science, Bangalore,
India -560012. \\
$^{2}$Department of Physics and Astronomy, University of Victoria, Victoria, BC V8P 1A1, Canada\\
$^{3}$Institute of Computational Science, centre for Theoretical Astrophysics and Cosmology, University 
of Zurich, \\ Winterthurerstrasse 190, 8057, Zurich, Switzerland
$^{4}$Institut d'Astrophysique de Paris-98\\ bis boulevard Arago-75014 Paris, France }
\maketitle

\begin{abstract}
Several arguments suggest that stochastic condensation of cold gas and its accretion onto the central supermassive black hole (SMBH)
is essential for active galactic nuclei (AGN) feedback to work in the most massive galaxies that lie at the centres of galaxy clusters.
Our 3-D hydrodynamic AGN jet-ICM (intracluster medium) simulations,  looking at the detailed angular momentum 
distribution of cold gas and its time variability for the first time, show that the angular momentum of the cold gas crossing $\lesssim 1$ kpc 
is essentially isotropic. With almost equal mass in clockwise and counter-clockwise orientations, we expect a cancellation of angular momentum
on roughly the dynamical time. This means that a compact accretion flow with a short viscous time ought to form, through which enough accretion 
power can be channeled into jet mechanical energy sufficiently quickly to prevent a cooling flow.
The inherent stochasticity, expected in feedback cycles driven by cold gas condensation, gives rise to a large variation in the cold gas mass at the centres of galaxy clusters, for similar cluster and SMBH masses, in agreement with the observations. Such correlations are expected to be much tighter
for the smoother hot/Bondi accretion. The weak correlation between cavity power and Bondi power obtained from our simulations also match observations.
\end{abstract}


\section{Introduction}
\label{sec:intro}

There is a general consensus that the intracluster medium (ICM) in low-entropy galaxy cluster cores is able to remain in rough \textit{global} thermal equilibrium because the powerful jets from the central AGN inject sufficient energy to compensate for radiative losses (for a review, see \citealt{mcn07,mcn12,fab12,sok16}). There are, however, a number of key issues associated with this ``radio feedback" schema that have yet to be firmly pinned down, one of which is: how is the central supermassive black hole (SMBH) able to transport sufficient energy to the cluster core ($\sim$ 10s of kpc) at a timescale shorter than the core cooling time 
($\lesssim 0.5$ Gyr). There are two scenarios proposed for AGN feedback: (1) hot/Bondi accretion which invokes spherical accretion with the mass flow rate set by the hot gas density and temperature on the cluster core scales (\citealt{bon52}); and (2) the competing cold mode accretion model, in which a fraction of the gas in the cluster cores condenses into cold clouds and filaments (when the ratio of the cooling time to the free-fall time [$t_{\rm cool}/t_{\rm ff}$] falls below a threshold value in the range 5-20; e.g., \citealt{mcc12,voi15}; but see \citealt{hog16}), which then decouple from surrounding hot intracluster medium (ICM), rain down upon the central SMBH, and power AGN feedback strong enough to prevent catastrophic cooling in the core.

Of the two, the Bondi accretion model, though more frequently invoked in the literature, has a number of shortcomings. For one, Bondi accretion is 
only applicable if the accretion flow is spherically symmetric and single phase, both conditions that are at odds with the multiphase gas with angular 
momentum observed in several cool cores (e.g., see \citealt{mac97,mcd10,mcn11,tre16}). Moreover, for a polytropic  hot flow $p\propto \rho^\gamma$, the mass 
accretion rate is given by (\citealt{shu92,fra02})  
\begin{eqnarray}
\nonumber
\dot{M}_B &=& 4 \pi \lambda(\gamma) G^2 M_{\rm BH}^2 \frac{\rho_\infty}{c_{s,\infty}^3} \\
&\approx& 0.0015  M_\odot {\rm yr}^{-1}~M_{\rm BH,9}^2 n_{\infty,0.1} T_{\infty, \rm keV}^{-3/2},
\label{eq:MdotB}
\end{eqnarray}
where $c_{s,\infty} \equiv [\gamma k_B T_{\infty}/\mu m_p]^{1/2}$ is the sound speed in the ambient medium far from the black hole; the mean particle mass $\mu=0.62$; $M_{\rm BH,9}$ is the black hole mass in units of $10^9 M_\odot$;  $n_{\infty,0.1}$ and $T_{\infty, \rm keV}$ are the ambient particle number density and temperature scaled to $0.1$ cm$^{-3}$ and 1 keV, respectively; and since $\lambda(\gamma)$ varies weakly with $\gamma$  
($\lambda$ varies from 1/4 for $\gamma=5/3$ to 1.1 for $\gamma=1$) we adopt $\gamma = 5/3$ for convenience.    In three nearby systems where both the ambient conditions governing the Bondi flow rate and the mass accretion rate onto the black hole can be deduced, the former is at least two orders of magnitude too large.
These nearby examples are Sgr A*, the Galactic centre BH [Bondi rate of $10^{-6}-10^{-5} M_\odot {\rm yr}^{-1}$ (\citealt{bag03}) vs. the  accretion rate at $\sim 10 R_{\rm Sch}$ of $\lesssim 10^{-7} M_\odot {\rm yr}^{-1}$ (\citealt{mar07} and references therein)]; NGC 3115 [Bondi rate of $2.2\times10^{-2} M_\odot {\rm yr}^{-1}$ vs. the  accretion rate onto the BH at least two orders of magnitude smaller (\citealt{won11,won14})]; and M87 [Bondi rate of $0.1-0.2$ $M_\odot {\rm yr}^{-1}$ vs. the  accretion rate onto the BH of $\lesssim 10^{-3} M_\odot {\rm yr}^{-1}$ (\citealt{rus15})].  

\citet{nem15} find that the median efficiency (defined as the ratio of the observed jet power and $\dot{M}_{\rm BH}c^2$; $\dot{M}_{\rm BH}$
 is the SMBH accretion rate) required to supply the cavity power of nearby radio galaxies is 300\%, assuming 
$\dot{M}_{\rm BH} \sim 0.01 \dot{M}_B$ typical of similar nearby hot accretion flows. This implies that hot accretion, with reasonable assumptions,
 is insufficient for powering most of the cavities observed in cluster cores.  
 At the extreme end of the spectrum, the estimated jet/cavity power in some of the galaxy clusters (e.g., MS0735, Cen A)  is found to be much larger than $0.1\dot{M}_Bc^2$; i.e., the Bondi mass flow rate is insufficient to power the observed X-ray cavities {\it even if all the hot gas in the Bondi flow is accreted by the SMBH} (\citealt{mcn11}).   And finally, the accretion rate in the Bondi  regime takes a long time to adjust to the conditions ($n,~T$) changing in the cluster cores ($\sim 1$ kpc) at the core cooling timescale $\sim$ few 100 Myr  (e.g., \citealt{sok09}). 

The cold mode feedback model sidesteps many of the challenges associated with the hot/Bondi model.   Very briefly (we refer the readers to \citealt{sha12}, \citealt{voi15}, and \citealt{pra15} [hereafter \citetalias{pra15}] for a more expansive discussion), the model is based on the realization that even if the ICM in the cluster core exists in rough global thermal balance, if the ratio of the cooling time and the gravitational free-fall time ($t_{\rm cool}/t_{\rm ff}$) falls below a critical value ($\sim 10$ or so; \citealt{sha12,gas12,voi15,pra15,li15,cho16}), the gas will become susceptible to local thermal instabilities and fragment, leading to the formation of a multiphase medium consisting of cold dense clouds condensing from the hot diffuse ICM itself. The clouds then fall towards central AGN, resulting in increased SMBH accretion and feedback that, in turn, quenches runaway cooling in the cluster core. 
The main question with this model is: will the infalling cold gas,  which has non-zero angular momentum, end up forming a viscosity-mediated standard accretion flow (SAF) in which the gas will flow inwards on a (long) viscous accretion timescale, or will the gas circularise sufficiently close to the SMBH and accrete on a timescale shorter than the core cooling time.

A particularly promising accretion mechanism, overcoming the angular momentum barrier by stochastically feeding cold gas clumps and fueling AGN activity, was  
invoked by \citet{piz05,piz10,hop06, nay07}. The AGN are stochastically fed by cloud clumps which cancel each others' angular momentum through 
inelastic collisions, leading to the formation of a small transient disk accreting onto the black hole (\citealt{nay07}). This turbulent accretion mechanism of cold gas clouds onto SMBH 
was later called cold chaotic accretion (\citealt{nay12}).  
Using isothermal simulations of idealised turbulent accretion flow over central 100 pc, \citet{hob11} show that dense gas accretes ballistically 
onto the central SMBH at a rate few order of magnitude larger than the case without turbulence.

\citet{gas13} carried out simulations of cluster cores (although only run for 40 Myr, a duration much shorter than the core cooling time) with idealised turbulent driving (which is fixed in time), and showed that the condensation 
of cold clumps from the hot ICM can boost the mass accretion rate on to the SMBH by $\sim 100$ compared to the Bondi accretion rate. These simulations were later generalised to include an initially rotating 
ICM (\citealt{gas15}) and cooling down to much lower temperatures allowing for three phases (hot, atomic and molecular; \citealt{gas17}), although still using idealised turbulence and running for much less 
than a core cooling time. The process of colliding cold gas clouds losing angular momentum and boosting SMBH accretion has been called chaotic cold accretion (CCA). Other terminology has also been used 
in past (ballistic, stochastic, forced accretion, etc.). In this paper we use a related but somewhat different term stochastic cold accretion 
(SCA) to highlight that a turbulent system such as the ICM should be described statistically (\citealt{mon71}).

While AGN jet-ICM simulations have been run successfully on cosmological timescales for some 
time now (e.g., \citealt{gas12,li15,pra15}), we investigate the detailed angular momentum distribution of cold gas and its implications on SMBH accretion in realistic jet-ICM simulations over cosmological time scales for the first time.
Here is a brief outline of the paper.
Section \ref{sec:sims} briefly presents our numerical setup. In section \ref{sec:ang_mom}, we calculate the circularization radius and viscous time of the cold  gas based on the standard accretion physics.
We show that the angular momentum distribution of cold gas crossing $\sim 1$ kpc is close to isotropic which will lead to angular momentum cancellation and the formation of an efficient accretion flow. We use the mass distribution of angular momentum to estimate the accretion rate onto the SMBH. 
In section \ref{sec:obs}, we compare some of the results from our simulations against key observational. In section \ref{sec:astro} we discuss the 
implications of our results, highlighting the need for better understanding of accretion. We summarise our paper in section \ref{sec:summary}.

\section{Numerical simulations}
\label{sec:sims}

\begin{table*}
\caption{List of runs}
\resizebox{1.0 \textwidth}{!}{%
\begin{tabular}{c c c c c c c c c c}
\hline
Run & $r_{\rm in}$  & $r_{\rm out}$ & $\epsilon^\dag$ & Run Time & $\dot{M}_{\rm in, cold/hot}$ & $M_{\rm cold}$ & $\dot{M}_B$ & $\dot{M}_{\rm BH, SAF}^\ddag$ 
& $\dot{M}_{\rm BH, SCA}^\ddag$  \\
       & (kpc) 	       & (kpc) & & (Gyr) &  ($M_{\odot}$yr$^{-1}$)    &  ($10^{11}~M_\odot$) & ($M_{\odot}$yr$^{-1}$) & ($M_{\odot}$yr$^{-1}$)  
       &  ($M_{\odot}$yr$^{-1}$)  \\
       & & & & & & & PL,~$\beta$& \\
\hline
NFW$^a$ &  1 & 200 & $6\times 10^{-5}$ & 4 & 18.1/3.5 & 1 & 0.2,~0.007 & 0.015 & 0.2 \\ 
NFW &  0.5 & 100 & $6 \times 10^{-5}$ & 5 & 10.7/0.9 & 5 & 0.54,~0.01 & 0.1 & 1.27 \\ 
NFW+BCG  & 0.5 & 500 & $5 \times 10^{-4}$ & 3 & 6.4/2.8 & 0.1 & 0.008,~0.003 & 0.001 & 0.03\\ 
\hline
\end{tabular}}
\label{tab:tab1}
\\
\begin{flushleft}
The resolution of all runs, done in spherical $(r_{\rm min} \leq r \leq r_{\rm max}, 0\leq \theta \leq \pi, 0 \leq \phi \leq 2\pi)$ coordinates, 
is $256\times128\times32$. Logarithmic grid is used in the $r-$ direction, and a uniform one in others. $\dot{M}_{\rm in, cold/hot}$ is 
the average cold/hot mass flow rate across $r_{\rm in}$; $\dot{M}_B$ is the estimate of average Bondi accretion rate using power-law (PL) and isothermal beta 
model ($\beta$) extrapolations; $M_{\rm cold}$ is the 
total cold ($T<0.1$ keV) gas mass in the simulation domain by the end.  $\dot{M}_{\rm in, cold/hot}$, $\dot{M}_{\rm BH,SAF}$, $\dot{M}_{\rm BH,SCA}$ \& $\dot{M}_{\rm B}$ are averaged from 1  Gyr till the end of the run. \\
$^a$ the fiducial run. \\
$^\ddag \dot{M}_{\rm BH, SAF/SCA}$ is the SMBH accretion rate calculated 
using Eq. \ref{eq:Mdot}/\ref{eq:Mdot_SCA}. \\
$^\dag$ jet efficiency relative to $\dot{M}_{\rm in}$ (see Eq. 6 in \citetalias{pra15}). \\
\end{flushleft}
\end{table*}

In this paper we discuss three 3-D AGN jet-ICM simulations carried out in spherical $(r,\theta,\phi)$ coordinates. 
The details of the numerical set up are given in \citetalias{pra15}. The three runs are: (i) an initially hydrostatic ICM in a fixed 
NFW potential (this is the fiducial run in \citetalias{pra15} 
with the halo mass of $7 \times 10^{14} M_\odot$); (ii) the same run but with the inner and outer radii reduced by half; (iii) and a NFW+BCG 
( the latter is modeled as a singular isothermal potential with circular velocity $V_c = 350$ km s$^{-1}$) potential run.
 The NFW+BCG run is carried out with {\tt PLUTO} (\citealt{mig07}) code, whereas the NFW runs are done with {\tt ZEUS-MP} (\citealt{hay06}).
 The primary purpose of the NFW+BCG runs is to facilitate a detailed analysis of how the addition of the BCG   
 potential impacts the various   aspects of the cold mode accretion model, a topic of considerable interest (c.f. \citealt{hog16,hog17}).  We will 
 present this analysis in a follow-up paper.   Most of the results and analysis in this paper are based on the NFW runs although we occasionally 
 call upon the NFW+BCG simulations to highlight that our key results are robust.

The initial entropy profile for both NFW and NFW+BCG runs are identical (see Eq. 7 in \citetalias{pra15}). 
The initial ICM density for the NFW+BCG run is almost twice of the NFW runs (which have identical initial density profiles). 
The different runs have different inner ($r_{\rm in}$) and outer ($r_{\rm out}$) 
radial boundaries (see Table \ref{tab:tab1}). The density and pressure at the outer radial boundary is fixed to their initial values. 
The total mass (including 
both hot and cold phases) crossing the inner boundary is calculated ($\dot{M}_{\rm in}$), and a fraction $\epsilon$ of $\dot{M}_{\rm in}c^2$ 
is put back into the ICM in the form of kinetic power ($P_{\rm jet}$) of AGN jets.

All our runs behave in a very similar manner. There are cooling and heating cycles, and the feedback driven heating events are less 
frequent for a higher feedback efficiency. Since the feedback efficiency, $\epsilon$, is ten times higher for the NFW+BCG run, the amount of cold gas and
accretion rate through the inner boundary are suppressed relative to NFW runs. 
The NFW run with $r_{\rm in}=0.5$ 
kpc uses the same efficiency parameter ($\epsilon$) as the fiducial run with $r_{\rm in}=1$ kpc. However, because of a smaller inner radius and non-zero angular momentum
of the condensing gas, 
the mass accretion rate across the inner boundary is lower as some of the gas circularises at $0.5~{\rm kpc} < r < 1$ kpc. Consequently, the feedback energy input is 
smaller and the run shows a 
higher accumulation of cold gas in the central regions 
(see $M_{\rm cold}$ in Table \ref{tab:tab1}). Density and temperature profiles, and other X-ray and jet properties are discussed in detail 
in \citetalias{pra15}, and will not be repeated in this paper. Here we focus on the angular momentum distribution of cold gas in our simulations and
the plausibility of stochastic accretion onto SMBHs (section \ref{sec:ang_mom}), and comparison of simulation results with observations (section \ref{sec:obs}).

\section{Angular momentum of cold gas \& SMBH accretion}
\label{sec:ang_mom}

Our simulations, like \citealt{gas12,li15,yan16}, use the total (dominated by cold gas) accretion rate at $\sim 1$ kpc to estimate the feedback energy deposited by jets. One of the major open problems with cold mode feedback is: how does the cold gas originating at $\gtrsim 1$ kpc lose angular momentum and get rapidly accreted onto the SMBH. This is needed for feedback to prevent a cooling flow. Although it is numerically formidable to resolve both the cluster core and accretion flow onto the SMBH, we attempt to estimate the accretion rate onto the SMBH based on the angular momentum distribution of cold gas within our simulation domain. While the large angular momentum gas forms a massive disk at $\gtrsim 1$ kpc, the smaller angular momentum gas crossing the inner boundary has stochastic angular momentum changing over short timescales. Rapid angular momentum cancellation in this stochastic cold gas can allow substantial cold gas to be channeled to the SMBH sufficiently fast to prevent a cooling flow (e.g., see \citealt{piz10}).

\subsection{Standard accretion estimates}

In this section we review the standard accretion physics that we apply later to estimate the accretion rate onto the SMBH (section \ref{sec:SAF}).  
For a given specific angular momentum ($l$), the circularization radius of the gas is given by
\begin{equation}
\label{eq:rcirc}
R_{\rm circ} \equiv \frac{l^2}{G M_{\rm enc}} \approx 0.24 {\rm~kpc}~ M_{\rm enc, 9}^{-1} l_{28}^2 ,
\end{equation} 
where $M_{\rm enc}$ is the mass enclosed (including only SMBH + Dark Matter contributions; i.e., ignoring gas mass) within the circularization radius. 
The viscous accretion time at the circularization radius can be expressed in terms of the 
specific angular momentum,
\begin{eqnarray}
\nonumber
t_{\rm visc}  &\approx& \frac{1}{\alpha \Omega (H/R)^2} \approx \frac{l^3}{\alpha G^2 M_{\rm enc}^2 (H/R)^2} \\
&\approx&~ 1.8 ~{\rm Gyr}~ \alpha_{0.1}^{-1} \left( \frac{H}{R} \right)_{0.1}^{-2} M_{\rm enc,9}^{-2} l_{28}^3,
\label{eq:tvisc}
\end{eqnarray} 
where $\Omega \equiv (GM_{\rm enc}/R^3)^{1/2}$ is the angular velocity, $\alpha$ is the viscosity parameter (\citealt{sha73}), and $H/R$ is the 
ratio of the disk height and radius. We scale the numerical values using $\alpha=0.1$. This is consistent with results from MHD 
simulations of magnetised accretion flows, which find that the effective value of $\alpha$ is 0.1 over the bulk of the flow (\citealt{haw01, bab13}).
Although the realistic $H/R$ for standard AGN disks is $\sim 10^{-3}$ (e.g., see Eq. 6 in \citealt{bab13}),  we scale our results to $H/R=0.1$ 
because the standard choice
gives a very small SMBH accretion rate. Moreover, stochastic accretion simulations show a thick cold flow
($H/R \gtrsim 0.1$; see section \ref{sec:SCA}) supported by turbulent motions (see Fig. 16 in \citealt{hob11, gas13}). Incidentally, radiatively inefficient accretion flows (RIAFs)
also have similar thick disks (\citealt{das13,yua14}).

\begin{figure}
	\includegraphics[scale=0.37]{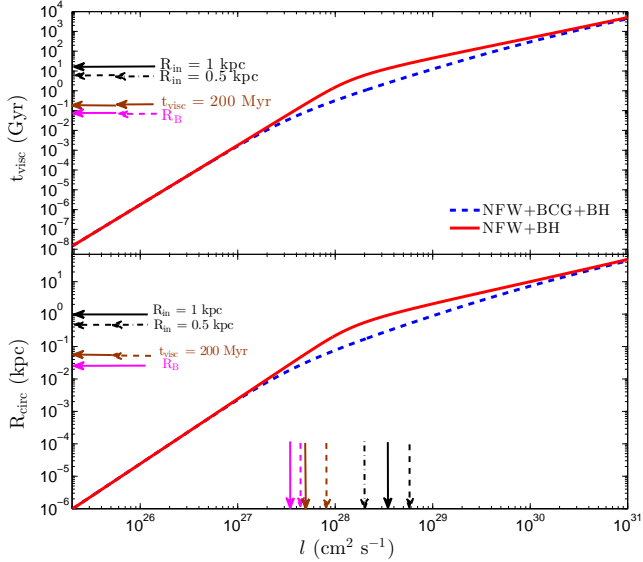}
	\caption{ Circularization radius (Eq. \ref{eq:rcirc}; bottom panel) and viscous time (Eq. \ref{eq:tvisc}; top panel) 
	as a function of specific angular momentum ($l$) for 
	NFW+SMBH and NFW+BCG+SMBH potentials. As expected,
	the SMBH dominates the enclosed mass within the Bondi radius. 
	Magenta arrows represent the Bondi radius, the corresponding $l$ 
	(Eq. \ref{eq:rcirc}), and $t_{\rm visc}$ (Eq. \ref{eq:tvisc}). Black arrows represent quantities corresponding 
	to the inner simulation radius of 0.5 kpc (dot-dashed line for NFW and dashed line for NFW+BCG) and 1 kpc 
	(solid line). Brown arrows show the viscous time of 0.2 Gyr.  These scales are useful
	to estimate SMBH accretion rates in section \ref{sec:acc_est}. }
	\label{fig:r_t_vs_l}
\end{figure}

Figure \ref{fig:r_t_vs_l} shows the viscous time (top panel) and circularization radius (bottom panel), as a function of the 
specific angular momentum,
for the two gravitational potentials (NFW and NFW+BCG) used in our simulations. Note that the inner radius of 
our computational domain 
is well beyond the Bondi radius, and the SMBH contribution to gravity is negligible everywhere in the computational 
domain ($r_{\rm in} \leq r \leq r_{\rm out}$). The slope of
$t_{\rm visc}$ and $R_{\rm circ}$ as a function of $l$ changes at the radius at which the enclosed mass is dominated by the extended
potential rather than the SMBH. The arrows in Figure \ref{fig:r_t_vs_l} mark important quantities such as Bondi radius, 
inner radius of the computational 
domain, and a viscous time of 0.2 Gyr.  Accretion flow with a viscous time more than the core cooling time 
(here taken to be 0.2 Gyr, a conservative estimate) would not be able to respond faster than the cooling rate. A faster response 
of the SMBH accretion rate is required to prevent a cooling flow.

\subsection{Stochastic cold accretion}

\label{sec:SCA}

\begin{figure*}
	\includegraphics[width=3.4truein, height=3.3truein]{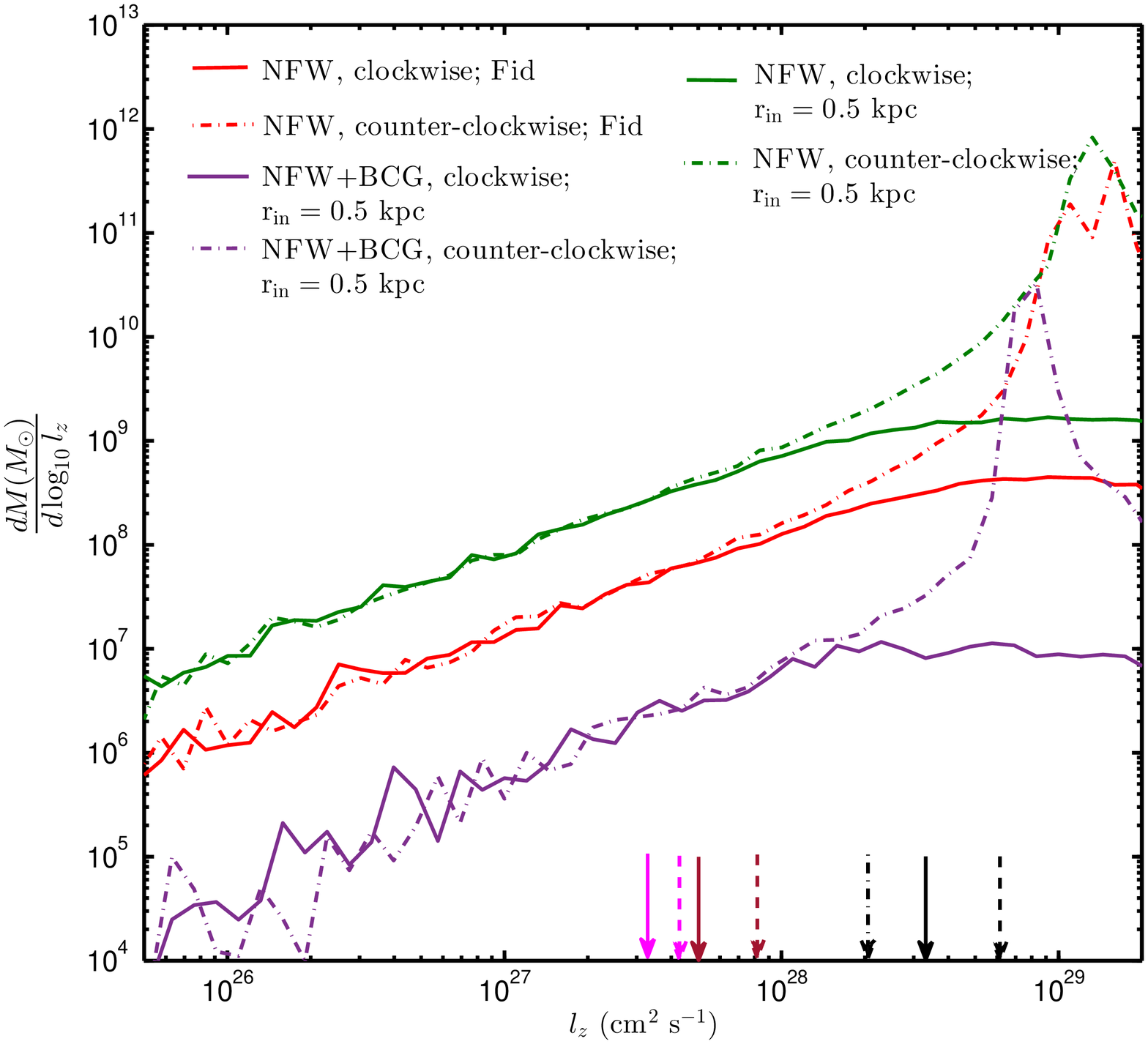}
	\includegraphics[width=3.5truein, height=3.3truein]{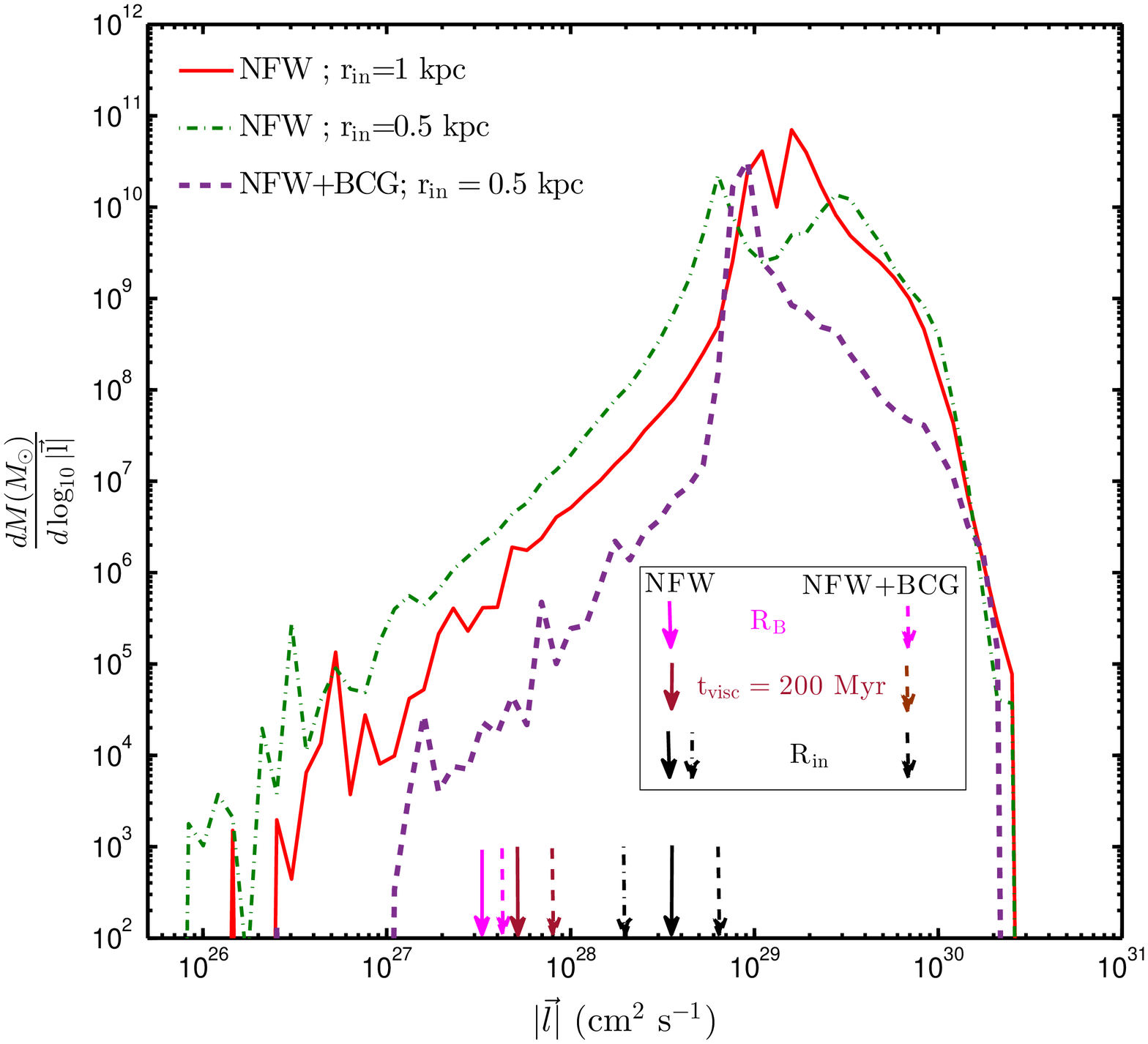}
	\caption{Time-averaged (between 2 and 3 Gyr) mass distribution of specific angular momentum 
	(left panel is with respect to $l_z$ and the right panel with respect to $|\bm{l}|$) in our simulations. 
	To calculate pdfs, equal-sized bins are used in $\log_{10} l$ space.
	Note that the slope of the mass distribution for $l_z$ and $|\bm{l}|<10^{28}$ cm$^2$s$^{-1}$ in all three simulations are very similar. 
	At smaller angular momenta ($|\bm{l}|<10^{28}$ cm$^2$s$^{-1}$), the mass distributions scale roughly as $dM/d l_z \propto l_z^0$ 
	and $dM/d|\bm{l}| \propto |\bm{l}|$. Notice the sudden rise in cold gas mass for specific angular momentum larger than the value 
	corresponding to the inner simulation domain ($l_{\rm in}$; marked by black arrows). This cold gas with large angular momentum 
	corresponds to the massive rotationally supported torus seen in our simulations. Note that in later figures we show averages 
	from 1 Gyr till the end of the simulation to get better statistics of time-averaged quantities (see Table \ref{tab:tab1}). We confirm that the slopes
	of the pdfs do not vary with the duration of averaging.
	}
	\label{fig:l_pdf}
\end{figure*}

In this section we discuss the angular momentum distribution of cold gas in our numerical simulations.
We show that the time-averaged angular momentum distribution of cold gas crossing our inner simulation boundary is stochastic, 
with almost equal mass going around in different directions. This implies that in the region of circularization, 
angular momentum cancellation will take place almost at the local dynamical time. Moreover, the cold gas with 
$l<10^{28}$ cm$^2$ s$^{-1}$ changes its mean angular momentum over a timescale shorter than the core cooling time; 
i.e., cold gas with small angular momentum is able to respond fast enough to close the feedback loop.

\subsubsection{Time-averaged angular momentum pdf}

To start with, we emphasise that when considering gas whose specific angular momentum ($\bm{l}$)  distribution is stochastic, 
it is essential to explicitly account for the vector nature of the specific angular momentum,
$\bm{l} \equiv \bm{r} \times \bm{v} \ (= -r v_\phi \bm{\hat{\theta}} + r v_\theta \bm{\hat{\phi}} = 
v_\phi R \bm{\hat{z}} - r[v_\phi \cos \theta \cos \phi + 
v_\theta \sin \phi] \bm{\hat{x}} + r[-v_\phi \cos \theta \sin \phi + v_\theta \cos \phi] \bm{\hat{y}} $; 
where $R/r$ is the cylindrical/spherical radius). In standard accretion disk 
literature, $l$ is generally taken to be $l_z$ since the gas is considered to be coplanar. 
However, in our simulations the  low angular momentum  cold gas distribution shows almost equal importance of all the components of $\bm{l}$.

 \begin{figure}
	\includegraphics[scale=0.37]{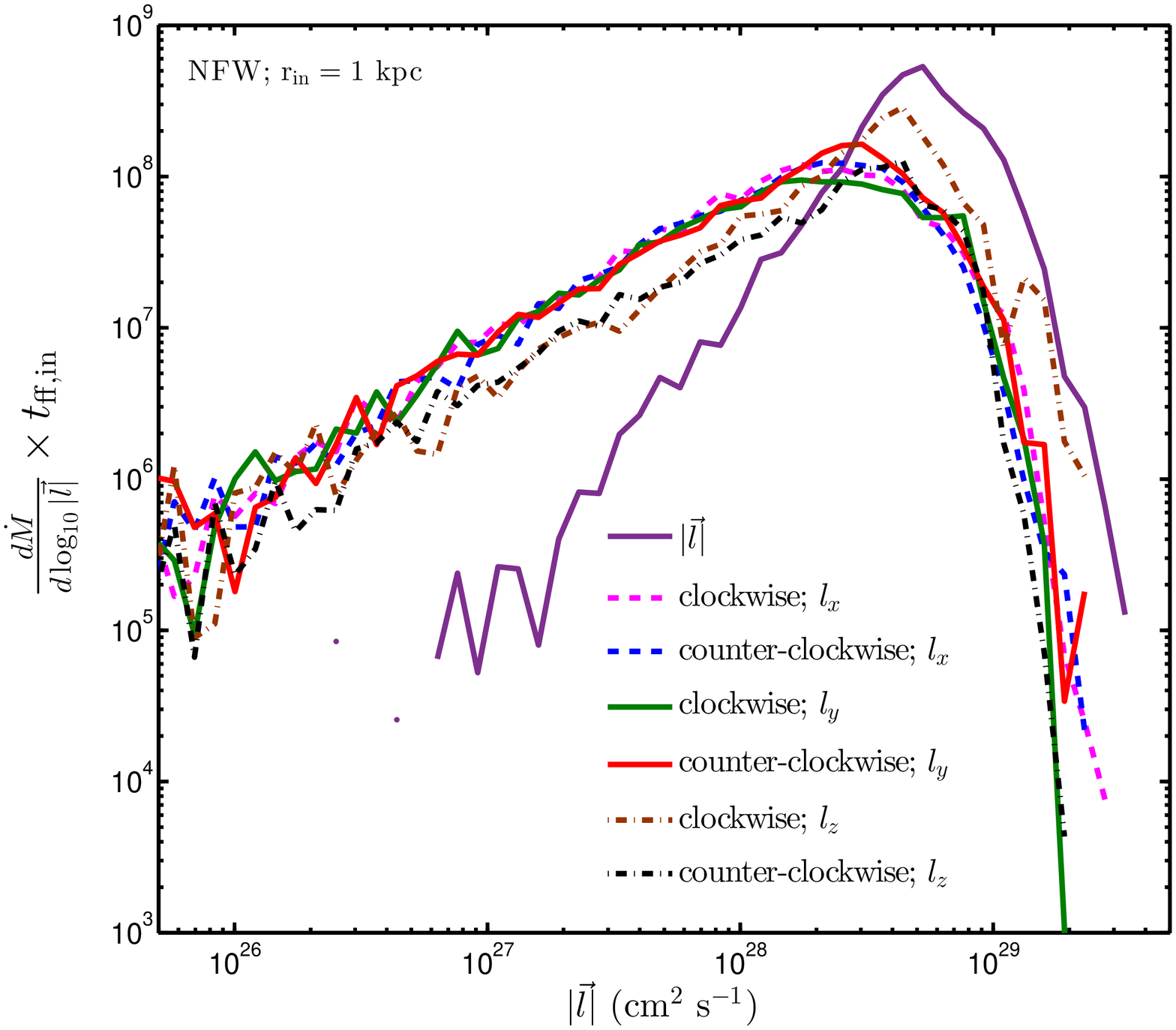}
	\caption{Time-averaged (between 1-4 Gyr) cold $\dot{M}$ (mass inflow rate of cold gas across the inner simulation boundary) distribution 
	(pdf) of $l_x,~l_y,~l_z$ and $|\bm{l}|$ for the fiducial
	NFW run with $r_{\rm in}=1$ kpc. The slopes are similar to the mass distributions in Fig. \ref{fig:l_pdf}. 
	The mass accretion rate is multiplied by $t_{\rm ff, in}$, the free-fall timescale at the inner boundary (15 Myr for the fiducial run).
	Not only are the $l_x$, $l_y$ and $l_z$ pdfs similar, the clockwise and
	counter-clockwise pdfs are also very similar. There is a slight bias in the $l_z$ distribution towards larger values of $l_z$ in the clockwise
	sense, which is a signature of the massive rotating torus  seen more prominently for the same run in Fig. \ref{fig:l_pdf}.  
	} 
	\label{fig:l_pdf_fid}
\end{figure}

To examine this, we consider the quantity $(dM/dl)\Delta l$, the incremental mass in the computational domain with angular momentum between $l$ and $l+\Delta l$
(we plot the mass distribution in a logarithmic bin $dM/d\log_{10} l =  [l/\log_{10} e]dM/dl$, as one can simply read off 
the mass from the figure). 
We can choose $l$ to be $|\bm{l}|$, or $l_x$, $l_y$ or $l_z$.  
The left panel of Figure \ref{fig:l_pdf} shows the time-averaged cold mass distribution as a function of $|l_z|$ (both clockwise and counter-clockwise 
rotations are shown) for our three runs. The right panel shows the time-averaged mass distribution as a function of the {\it total} specific 
angular momentum $|\bm{l}|$. 
Comparing the $|\bm{l}|$ and $|l_z|$ distributions, we find a dramatic lack of
cold gas with small $|\bm{l}|$ but a non-negligible mass of cold gas with very low $|l_z|$. This gas with low $|l_z|$ has large $|l_x|$ and $|l_y|$ components, and therefore is not expected to circularise in the x-y plane. 

Focussing on gas with $l<l_{\rm in}$ (corresponding to the circularization radius in Eq. \ref{eq:rcirc} equal to the inner radial boundary $r_{\rm in}$), where 
$l_{\rm in} \sim 10^{29}$ cm$^2$s$^{-1}$, we find that in all our simulations the low angular momentum gas ($l<l_{\rm in}$) has roughly 
similar distribution in $l_z$ and $|\bm{l}|$, with $dM/dl_z \propto l_z^0$ and $dM/d|\bm{l}| \propto |\bm{l}|$.
\footnote{While we have not investigated the origin of these power laws, we note that the power law exponents are analogous to the Maxwellian distribution at low velocities (compared to thermal); namely the 1-D distribution $f \propto v_x^0$ 
(with respect to the velocity along a fixed axis) is shallower compared to the distribution with respect to  the absolute value of the total velocity $f \propto v^2$.  
}  
In fact, both the clockwise and counter-clockwise components of angular momentum 
distribution have very similar profiles for $l \lesssim l_{\rm in}$. Since the mass probability density function (pdf) in Fig. \ref{fig:l_pdf} does not peak as abruptly 
for simulations with lower $l_{\rm in}$ (and smaller $r_{\rm in}$), we assert that the low-$l$ mass pdfs are robust (in the sense that the low-$l$ pdf will be similar
if $r_{\rm in}$ is reduced further).

Also, as apparent in Figure \ref{fig:l_pdf}, there is an excess mass at specific angular momenta larger than $l_{\rm in}$. This matter with large angular momentum 
(primarily $l_z$) circularises within the computational domain (see the cold torus in Fig. 3 of \citetalias{pra15}) and does not cross the inner boundary, 
unlike the low angular momentum cold gas.

\begin{figure*}
	\includegraphics[scale=0.43]{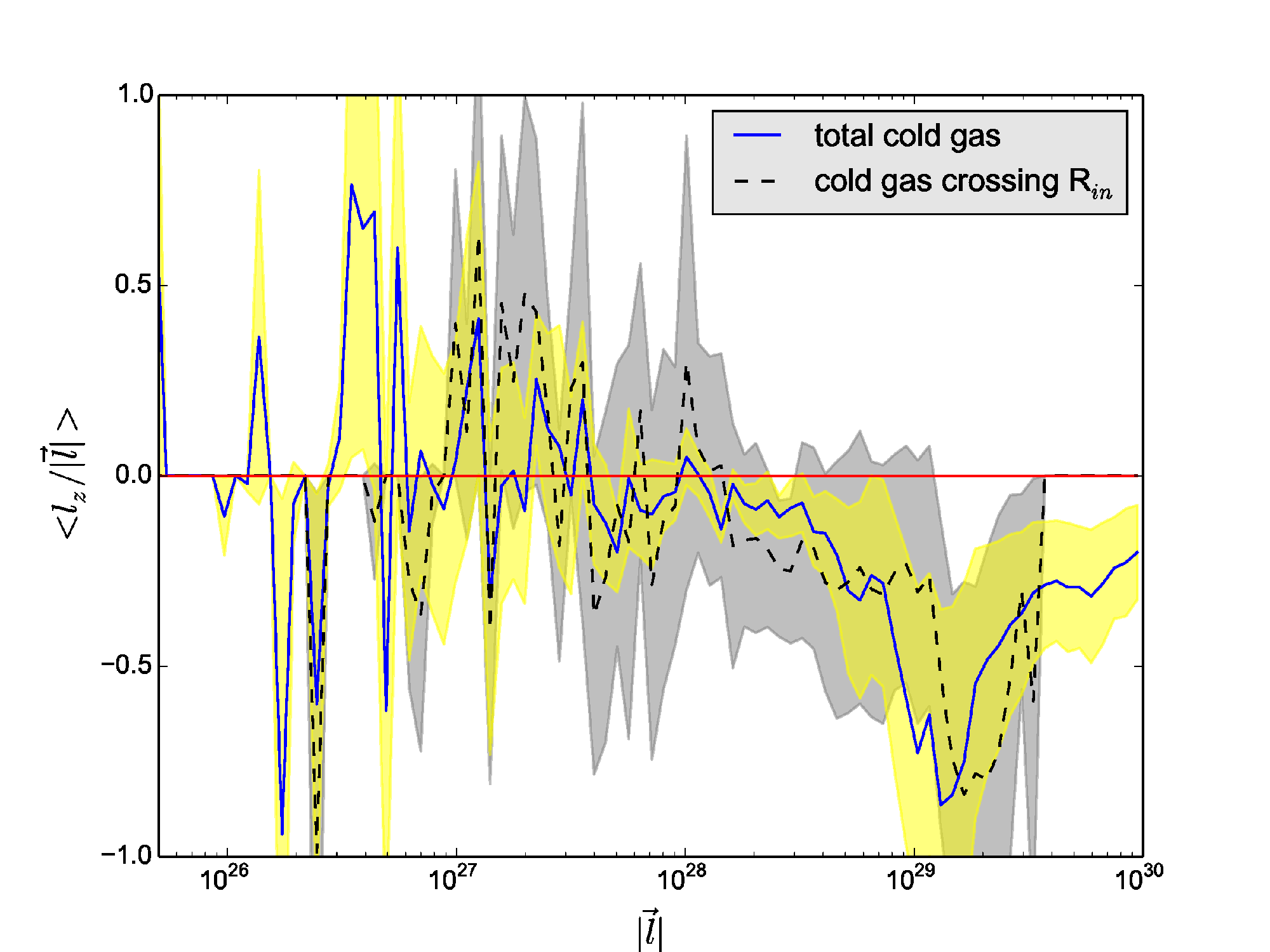}
	\includegraphics[scale=0.43]{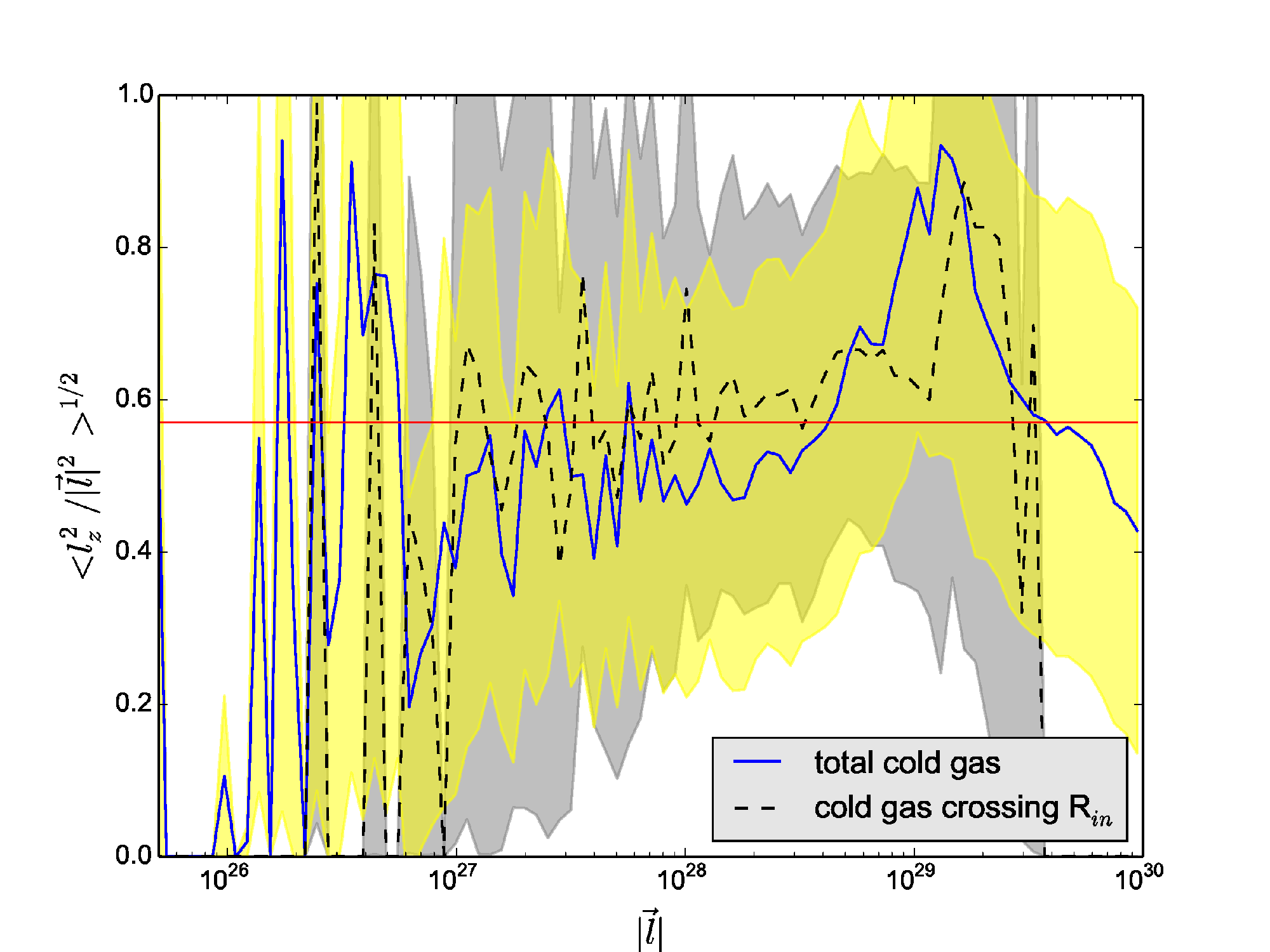}
	\caption{ The orientation of cold gas from 1-4 Gyr in the fiducial simulation characterised by $\theta$, the angle between the 
	$z-$ axis and the specific angular momentum vector ($\bm{l}$). The left panel shows the distribution of $\cos \theta = \langle l_z/|\bm{l}| \rangle$ and the
	right panel shows  rms $\cos \theta = (\langle l_z^2/|\bm{l}|^2 \rangle)^{1/2}$ ($\langle \rangle$ represents mass-weighted averaging for each
	$|\bm{l}|$ bin). 
	For perfectly isotropic distribution, the mean value of $\cos \theta$ 
	should be 0 and that of rms $\cos \theta$ should be $1/\sqrt{3} = 0.58$ (these values are marked by the horizontal solid red lines). 
The blue solid line (black dashed line) is the mass-weighted, time-averaged orientation as a function of $|\bm{l}|$ for the cold gas within 
the domain (crossing the inner boundary). 
The shaded yellow (grey) region represents the $1\sigma$ spread around the mean of cold gas within the computational domain (crossing the inner
boundary). Unlike the cold gas within the simulation box, there is no cold gas with 
large angular momentum crossing the inner boundary (also seen by comparing Figs. \ref{fig:l_pdf} \& \ref{fig:l_pdf_fid}). Note that the time-averaged 
angular momentum distribution becomes isotropic for $|\bm{l}| \lesssim 10^{28}$ cm$^2$s$^{-1}$.}
\label{fig:l_scatter}
\end{figure*}

\subsubsection{Angular momentum pdf of mass crossing $r_{\rm in}$}

Apart from the angular momentum pdf of cold gas mass {\it within the computational domain}, it is instructive to consider the pdf of the cold gas
{\it crossing the inner boundary}, which is more relevant for SMBH accretion estimates.
Figure \ref{fig:l_pdf_fid} shows the time-averaged angular momentum pdf ($d\dot{M}/d\log_{10} l$) with respect to $l_x,~l_y,~l_z,$ and $|\bm{l}|$ of 
cold gas accreting through the inner boundary of our fiducial run. The pdfs relative to all three components ($l_x$, $l_y$, $l_z$) are similar, including the clockwise and counter-clockwise components, suggesting that the cold gas crossing the inner boundary is roughly isotropic. 
The $d\dot{M}/d\log_{10} l$ pdf with respect to $|\bm{l}|$ is truncated at higher $|\bm{l}|$, cutting off sharply at $|\bm{l}| \approx 4 \times 10^{28}$ cm$^2$ s$^{-1}$, which corresponds to the specific angular momentum at the inner 
boundary ($l_{\rm in}$; see Fig. \ref{fig:l_pdf}); the gas with larger angular momentum cannot overcome the centrifugal barrier and fall in through the inner boundary.
The accretion rate pdfs in Figure \ref{fig:l_pdf_fid} and the mass pdfs in Figure \ref{fig:l_pdf} have similar slopes at low $l$ with respect to $l_x,l_y,l_z$ and 
$|\bm{l}|$.

The left panel of Figure \ref{fig:l_scatter} represents the rotational direction  ($\cos \theta = l_z/|\bm{l}|$) of cold gas at all times in the fiducial NFW run. The right panel shows the corresponding rms $\cos \theta = \langle l_z^2/|\bm{l} |^2 \rangle^{1/2}$ ($\langle \rangle$ stands for mass-weighted averaging). 
The blue solid (black dashed) line is the time-averaged, mass-weighted orientation of cold gas {\it within} ({\it crossing}) the computational domain. 
 The grey shaded region shows the $1\sigma$ scatter of cold gas  {\it within} the computational domain while the yellow shaded region represents 
the  $1\sigma$ scatter of the cold gas {\it crossing} the inner boundary.
The cold gas with $|\bm{l}| \lesssim 10^{28}$ cm$^2$s$^{-1}$ has equal scatter around the value expected for the mean and rms angle between $\bm{l}$ and the $z-$ axis for isotropic distribution ($\langle \cos \theta \rangle=0$ and 
$\langle \cos^2\theta \rangle^{1/2}=1/\sqrt{3} \approx 0.58 $), implying that the time-averaged angular momentum distribution of the cold gas with small angular momentum within the simulation domain 
and crossing the  inner boundary is almost
isotropic. The larger angular momentum gas, on the other hand, has a clockwise bias as also seen in Figures \ref{fig:l_pdf} \& \ref{fig:l_pdf_fid}.

\begin{figure}
	\includegraphics[scale=0.38]{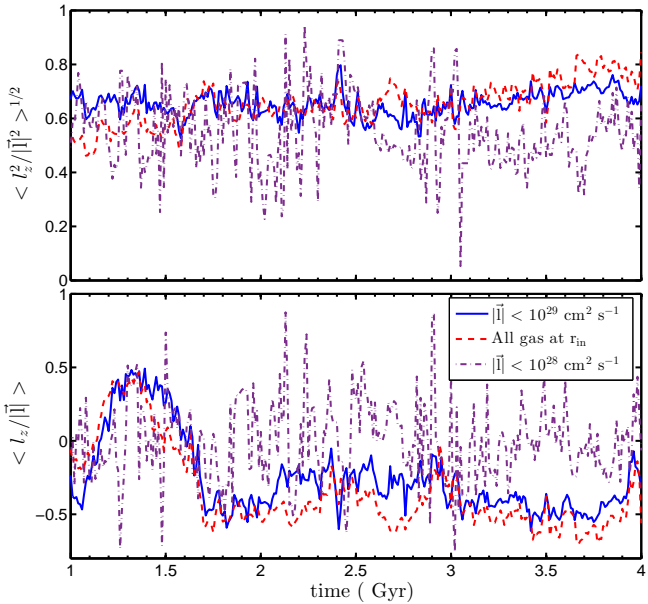}
	\caption{Average mass-weighted $\langle {l_z}/{|\bm{l}|} \rangle$ (bottom panel) and rms $\langle l_z^2/|\bm{l}|^2\rangle^{1/2}$ (top panel) of cold 
	gas crossing the inner boundary as a function of time for the fiducial NFW run.
	The dashed red lines are for all the cold gas crossing the inner boundary.
	The purple dot-dashed lines with stars show the average orientation of cold gas with $|\bm{l}| < 10^{28}$ cm$^2$s$^{-1}$ and 
	solid blue lines are for  $|\bm{l}|$ $<10^{29}$ cm$^2$s$^{-1}$. The orientation of cold gas with $|\bm{l}|<10^{28}$ 
	cm$^2$s$^{-1}$ changes on a short timescale compared to the other two cases. 
	}
	\label{fig:l_pdf_time}
\end{figure}

\subsubsection{Time dependence of angular momentum pdf}

Now that we have shown that the {\it time-averaged} angular momentum distribution of cold gas with $|\bm{l}| \lesssim 10^{28}$ erg s$^{-1}$ is 
roughly isotropic, we want to study its time dependence. In particular, stochastic accretion of cold gas is a viable solution of the cooling flow
problem only if the low angular momentum cold gas changes direction on a time scale shorter than the core cooling time 
(taken to be 0.2 Gyr  as a conservative estimate; for a longer cooling time, more cold gas can power SMBH accretion). 
If the angular momentum variability time scale is shorter, the cold gas can undergo angular momentum cancellation (and hence accretion) faster than the core cooling time.

Figure \ref{fig:l_pdf_time} shows the orientation of the average mass-weighted angular momentum distribution of
cold gas crossing the inner radial boundary ($\langle l_z/|\bm{l}| \rangle$ and rms $\langle l_z^2/|\bm{l}|^2 \rangle^{1/2}$, same quantities as 
in Fig. \ref{fig:l_scatter}) as a function of time. 
The orientation of the total cold gas, and $|\bm{l}|<10^{29}$ cm$^2$s$^{-1}$ and $|\bm{l}|<10^{28}$  cm$^2$s$^{-1}$ cold gas 
 are shown. The $l_z/|\bm{l}|$ ratios of the total cold gas and with $|\bm{l}|<10^{29}$ cm$^2$s$^{-1}$ 
show that this gas only changes its orientation (as measured by $\cos \theta$ crossing zero) till only 1.5 Gyr, and attain a fixed sense of rotation thereafter. 
Even though this cold gas (dominated by large angular momentum gas) shows small fluctuations on short time scales (< 30 Myr), it still maintains its orientation 
for the rest of the simulation time. Time of 1.5 Gyr coincides with the formation of the massive cold torus in the fiducial simulation 
(see Fig. 3 in \citetalias{pra15}). The rms $l_z^2/|\bm{l}|^2$ value for this cold gas remains well above 0.58, suggesting that most of it settles down in a 
disk-like structure. Note that both $\langle l_z/|\bm{l}| \rangle$ and rms $\langle l_z^2/|\bm{l}|^2 \rangle^{1/2}$ (see the purple dashed lines with markers
 in Fig. \ref{fig:l_pdf_time}) are much more variable for low angular 
momentum cold gas ($|\bm{l}| < 10^{28}$ cm$^2$s$^{-1}$) as compared to the total cold gas. The mean angle between $\bm{l}$ and the $z-$ axis 
($\langle \cos \theta \rangle$) fluctuates around zero and the rms $\langle \cos^2 \theta \rangle$ fluctuates around the isotropic value (0.58) on a timescale shorter 
than 0.2 Gyr. Therefore, only the low angular momentum gas ($|\bm{l}|<10^{28}$ cm$^2$s$^{-1}$) is expected to participate in feedback heating of the ICM.  
  
Results from our AGN jet-ICM simulations (Figure \ref{fig:l_pdf_time}) show that cold gas accretion at $\lesssim 1$ kpc is stochastic, with the angular momentum of cold gas with $|\bm{l}|<10^{28}$ 
cm$^2$s$^{-1}$ changing on time scales shorter than the core cooling time. The mixing and inelastic cloud-cloud collisions of these randomly oriented cold gas clouds near the circularization radius will result in angular momentum cancellation. After losing angular momentum these clouds will fall ballistically to the centre, where they ought to establish a compact turbulent accretion flow with $H/R \gtrsim 0.1$ (see \citealt{hob11}) and viscous time shorter than 0.2 Gyr, the cooling time of the cluster core. From our simulation results, a specific angular momentum $ \lesssim 10^{28}$ cm$^2$s$^{-1}$ can be taken as the limit for stochastic cold accretion (see Figs. \ref{fig:l_scatter} \& \ref{fig:l_pdf_time}).   

\subsection{SMBH accretion estimates}
\label{sec:acc_est}
In this section we estimate the mass accretion rate onto the central SMBH, first using the standard steady accretion flow model and then using the 
more plausible stochastic cold gas accretion. For the standard flow model estimate, we assume that cold gas forms a thick extended disk before 
accreting onto the central SMBH on the local viscous time at the circularization radius. For stochastic cold accretion, the angular momentum of 
the randomly oriented gas clouds cancel each other and the clouds settle down in a compact accretion flow very close to central SMBH. In this scenario accretion 
happens at the dynamical time (roughly the timescale for angular momentum cancellation) because instead of relying on viscosity to transport out angular 
momentum, accretion is powered by angular momentum cancellation among infalling clouds with random orientations.  

Typical efficiency required with respect to the mass accretion rate $\dot{M}_{\rm in}$ at $\sim 1$ kpc to suppress the cooling rate to at least $10$\% of the pure 
cooling flow value, $\dot{M}_{\rm cf}$, is $\sim 10^{-4}$. 
Our cluster mass is $M_{200} = 7\times10^{14} M_\odot$; this $\epsilon \equiv P_{\rm jet}/\dot{M}_{\rm in} c^2$ is expected to be smaller for lower mass halos (see Fig. 8 in \citetalias{pra15}), where $P_{\rm jet}$ is the jet power.
This means that, assuming a SMBH mechanical efficiency of $10\%$ ($P_{\rm jet} = 0.1 \dot{M}_{\rm BH} c^2$), $\dot{M}_{\rm BH} \gtrsim 10^{-3} \dot{M}_{\rm 1kpc}$ is required to sufficiently suppress a cooling flow; i.e., at least 0.1\% of the 
cold gas infalling at 1 kpc must be 
accreted by the SMBH on a timescale shorter than the core cooling time. Thus, for our fiducial run (for which $\dot{M}_{\rm in} \approx 20 M_\odot {\rm yr}^{-1}$; 
see Table \ref{tab:tab1}) average $\dot{M}_{\rm BH} \gtrsim 0.02 M_\odot {\rm yr}^{-1}$ is required to suppress a cooling flow.

\subsubsection{Viscosity-mediated standard accretion flow estimate}
\label{sec:SAF}

The time scale on which the gas crossing the inner boundary reaches the SMBH is determined by 
the nature of the accretion flow. If the accreting cold gas has zero angular momentum it falls at the 
free-fall rate, but in presence of angular momentum, the time for matter to cross the inner boundary and reach the SMBH is the
sum of the local free-fall time at the inner radius ($t_{\rm in}$) and the viscous timescale at the circularization radius 
($t_{\rm visc}$). Using these considerations, we can estimate the viscosity-mediated standard accretion flow (SAF) rate onto the 
SMBH, using the mass distribution of the cold gas crossing the inner boundary ($d\dot{M}/dl$; here $l=|\bm{l}|$), as
\begin{equation}
\label{eq:Mdot}
\dot{M}_{\rm BH, SAF} \approx \int_{l_{\rm ISCO}}^{l_{\rm max}} dl \frac{d\dot{M}}{dl} \frac{t_{\rm ff, in}}{[t_{\rm ff,in} + t_{\rm visc}(l)]},
\end{equation}
where  
$l_{\rm max}$ is the maximum angular momentum of the cold gas that can contribute to accretion and $l_{\rm ISCO}$ is the angular momentum 
corresponding to the innermost stable circular orbit (ISCO), 
which for a Schwarzschild blackhole is, $r_{\rm ISCO}=6 G M_{\rm BH}/c^2 \approx 2.9\times10^{-4} M_{\rm BH,9}$ pc. An appropriate upper angular momentum cut-off ($l_{\rm max}$) is the angular momentum for which the 
viscous accretion time equals 0.2 Gyr, the core cooling time (we show later in this section that $\dot{M}_{\rm BH, SAF}$ is rather insensitive 
to $l_{\rm max}$).  
 From Eq. \ref{eq:tvisc}, the specific angular momentum for a viscous time of 0.2 Gyr is 
$0.48 \times 10^{28}$ cm$^2$s$^{-1}$. Gas with a larger angular momentum accretes on a time scale longer than the 
core cooling time and hence cannot stop a cooling flow. Moreover, matter with angular momentum smaller than $l_{\rm ISCO} \equiv \sqrt{12} 
GM_{\rm BH}/c \approx 1.5 \times 10^{25} M_{\rm BH,9}$ cm$^2$ s$^{-1}$ (assuming a non-rotating BH) falls directly into the SMBH
 with no way to extract its gravitational potential energy.

The low angular momentum distribution of the gas crossing the inner simulation boundary (with respect to $|\bm{l}|$; see Fig. \ref{fig:l_pdf_fid}) 
can be roughly approximated as $d \dot{M}(l)/dl_{28} \sim 0.56 M_\odot {\rm yr}^{-1} l_{28}$ ($l_{28}=l/10^{28} {\rm cm^2 s^{-1}}$). Eq. \ref{eq:Mdot},
with this approximation, can be written as
\begin{equation}
\label{eq:Mdot_nondim}
\dot{M}_{\rm BH, SAF} \approx 0.56 M_\odot {\rm yr}^{-1} \int_{l_{\rm ISCO,28}}^{l_{\rm max,28}} dl_{28} \frac{l_{28}}{[1 + t_{\rm visc}(l_{28})/t_{\rm ff,in}]}.
\end{equation}
From Eq. \ref{eq:tvisc}, we can write 
\begin{equation}
\frac{t_{\rm visc}(l_{28})}{t_{\rm ff,in}} = 120 \left (\frac{t_{\rm ff, in}}{15 {\rm Myr}} \right )^{-1} \alpha_{0.1}^{-1} \left (\frac{H}{R}\right )_{0.1}^{-2} M_{\rm BH, 9}^{-2} l_{28}^3.
\end{equation}
Taking $l_{\rm max,28}=0.48$, which corresponds to a viscous time equal to the core cooling time (Eq. \ref{eq:tvisc}) for the fiducial parameters,\footnote{Eq. 
\ref{eq:mdot_simp} is not sensitive to the upper cut-off as $\dot{M}_{\rm BH,SAF} \propto l_{28}^{-1}$.}
\begin{equation}
\label{eq:mdot_simp}
\dot{M}_{\rm BH,SAF} = 0.56 \int_{1.5 \times 10^{-3}}^{0.48} dl_{28} \frac{l_{28}}{(1+120 l_{28}^3)} \approx 0.018 M_\odot {\rm yr}^{-1}.
\end{equation}
This value is uncomfortably close to the minimum value required to prevent a cooling flow ($\epsilon \dot{M}_{\rm in}/0.1 \sim  0.02 M_\odot~{\rm yr}^{-1}$).
Moreover, we have used a large 
$H/R=0.1$ instead of the standard thin AGN disk value of $H/R\sim 10^{-3}$. Using $H/R=10^{-3}$ gives $l_{\rm max, 28}=0.022$ and 
$\dot{M}_{\rm BH, SAF} = 1.3 \times 10^{-4} M_\odot {\rm yr}^{-1}$, way too small compared to the required accretion rate.  The value of
the $\alpha$ viscosity parameter is uncertain, with the variability observations giving an order of magnitude larger value ($\gtrsim 0.1$) compared to 
MHD simulations without net flux (see \citealt{kin07} and references therein). A smaller $\alpha$ makes the case for stochastic cold accretion (SCA) even stronger.
We list the actual value of the integral in Eq. \ref{eq:Mdot}  (rather than making a power-law approximation) in Table \ref{tab:tab1} for $H/R=0.1$.

\subsubsection{Stochastic cold accretion (SCA) estimate}

Accretion for the stochastic cold gas (Figs. \ref{fig:l_scatter} \& \ref{fig:l_pdf_time} show that $|\bm{l}| < 10^{28}$ cm$^2$s$^{-1}$ cold 
gas is stochastic in space and time over 0.2 Gyr) is expected to proceed at almost the local free-fall time till the gas settles in a thick disk close to the
ISCO where the viscous time is very short compared to the free fall at the inner boundary of our simulations, $t_{\rm ff, in}$. From Eq. \ref{eq:tvisc}, 
the viscous accretion time for $l=2 \times 10^{27}$ cm$^2$ s$^{-1}$ is 14 Myr. We assume that the stochastic cold gas with initial $l \lesssim 10^{28}$ 
cm$^2$ s$^{-1}$ cancels its angular momentum to $\lesssim  2\times10^{27}$ cm$^2$ s$^{-1}$ on a timescale shorter than 15 Myr (the free-fall time at
the inner boundary). 
The stochastic cold accretion (SCA) rate, therefore, can be approximated as
\begin{equation}
\label{eq:Mdot_SCA}
\dot{M}_{\rm BH,SCA} \approx \int_{l_{\rm ISCO}}^{l_{\rm max}} dl \frac{d \dot{M}}{dl},
\end{equation}
where $l=|\bm{l}|$. This estimate assumes that most of the mass accreting onto the SMBH originates as cold gas beyond $\sim 1$ kpc. Not much
gas condenses within 1 kpc because at most times $t_{\rm cool}/t_{\rm ff}>10$ at these radii. Moreover, the available hot gas mass (out of which cold gas can condense)
decreases inward. 
We take $l_{\rm max}=10^{28}$ cm$^{2}$ s$^{-1}$ as the upper cut-off of the integral in Eq. \ref{eq:Mdot_SCA} because cold gas is stochastic only for angular momentum
lower than this.

The integral in Eq. \ref{eq:Mdot_SCA} can be approximated as $\dot{M}_{\rm BH,SCA} \approx 0.2 M_\odot {\rm yr}^{-1} l_{\rm max, 28}^2$ (see Table \ref{tab:tab1} 
for the exact value) for our fiducial run for which $d \dot{M}/dl_{28} \approx 0.56 M_\odot {\rm yr}^{-1} l_{28}$. 
Note that the SCA estimate is independent of disk parameters ($\alpha$ \& $H/R$) as we assume that accretion in the SCA regime happens 
at the local free-fall time (which is shorter than the free-fall time at the inner boundary). Unlike viscosity-mediated standard accretion flow (SAF) integral (Eq. \ref{eq:Mdot}) which is rather insensitive to $l_{\rm max}$, 
SCA integral (Eq. \ref{eq:Mdot_SCA}) increases quadratically with $l_{\rm max}$.
Note that the SMBH mass accretion rate in the SCA regime is comfortably larger than
 the required $\dot{M}_{\rm BH} \sim 0.02 M_\odot {\rm yr}^{-1}$ to quench a cooling flow. In fact, we can also comfortably accommodate the reduction
 of $\dot{M}_{\rm BH}$ in the RIAF regime due to outflows.
 
 \begin{figure}
	\includegraphics[scale=0.4]{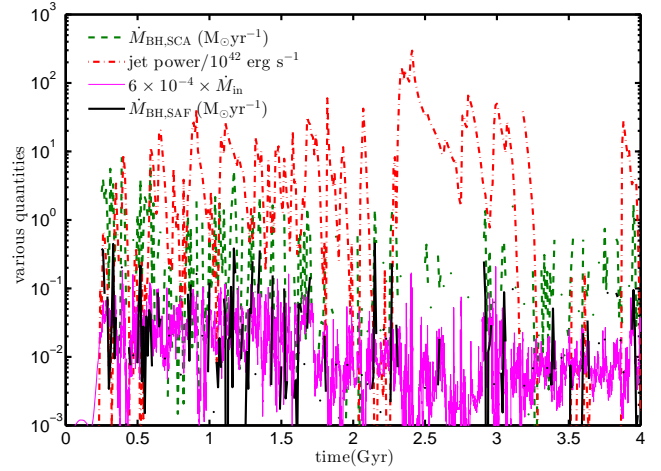}
	\caption{Mass accretion rate estimates as a function of time based on standard accretion flow (SAF in black thick solid line; Eq. \ref{eq:Mdot}) and 
	stochastic cold accretion (SCA in green dashed line; Eq. \ref{eq:Mdot_SCA}) for our fiducial run. Also shown in magenta solid lines is the expected
	mass accretion rate onto the SMBH assuming that the SMBH accretion rate is $\epsilon \dot{M}_{\rm in}/0.1 = 6 \times 10^{-4} \dot{M}_{\rm in}$. 
	Red dot-dashed line is the jet power 
	(see section \ref{sec:Bondi} on how we calculate the jet/cavity power) as a function of time. All quantities show large variations because of 
	heating/cooling cycles driven by condensation/overheating. The accretion rate at $r_{\rm in}$ is less time variable compared to $\dot{M}_{\rm SCA}$
	and $\dot{M}_{\rm SAF}$ but overall trends are similar. }
	\label{fig:mdot_vs_time}
\end{figure}

Figure \ref{fig:mdot_vs_time} shows the jet power, and the mass accretion rates estimated from standard accretion flow (SAF; Eq. \ref{eq:Mdot} 
with $H/R=0.1$) 
and stochastic cold accretion (SCA; Eq \ref{eq:Mdot_SCA}) as a function of time. As expected, the stochastic cold accretion (SCA) mass accretion rate is higher on average 
as compared to viscosity-mediated standard accretion flow (SAF).
Also $\dot{M}_{\rm SCA}$, in comparison to $\dot{M}_{\rm SAF}$, is variable on a shorter time. Notice the spikes in $\dot{M}_{\rm SCA}$ in the cooling phases
of the core and the lack of stochastic accretion when jets overheat the core. Similar trends are also seen for $\dot{M}_{\rm in}$; i.e.,  all accretion rates are higher in the 
cooling phases when jet power is small and suppressed in the overheating phases. 
We discuss the application of stochastic 
cold accretion estimate based on the angular momentum distribution of cold gas as the basis of a more robust feedback prescription 
in section \ref{sec:improv}.

\subsection{Improving cold feedback prescription}
\label{sec:improv}

Table \ref{tab:tab1} lists important time-averaged quantities from our numerical simulations. The accretion rate estimates based on the standard accretion flow (SAF; Eq. \ref{eq:Mdot} with $H/R=0.1$) and stochastic cold accretion (SCA; Eq. \ref{eq:Mdot_SCA} with $l_{\rm max} =4.8 \times 10^{27}$ cm$^2$s$^{-1}$) 
are included, as is the average rate of mass crossing the inner boundary. The NFW run with $r_{\rm in}=0.5$ kpc but the same $\epsilon$ as the fiducial NFW run (with $r_{\rm in}=1$ kpc) shows that 
the average mass
accretion (hot+cold) through the inner boundary is reduced by a factor of $\approx 2$ (11.6 $M_\odot {\rm yr}^{-1}$ as opposed to 21.6 $M_\odot {\rm yr}^{-1}$ 
for the fiducial run) but the cold gas mass accumulating within the computational domain (dominated by the high angular momentum gas; e.g., see Fig. \ref{fig:l_pdf}) by the end of the simulation is about five times higher. In fact, 
the average cold gas mass accumulation rate for the $r_{\rm in}=0.5$ kpc run is $5\times 10^{11} M_\odot / 5 {\rm Gyr} \sim 100 M_\odot {\rm yr}^{-1}$, 
only a factor of 2 lower than the pure cooling flow rate. A lot of cold gas crossing 1 kpc circularises before crossing the inner boundary at 0.5 kpc 
(compare the angular momentum pdfs and $l_{\rm in}$ for different runs in the right 
panel of Fig. \ref{fig:l_pdf}), and the mass accretion
rate at $r_{\rm in}=0.5$ kpc ($\dot{M}_{\rm in}$) is much smaller than the average mass cooling/deposition rate. 
Importantly, a smaller $\dot{M}_{\rm in}$ for a fixed $\epsilon$ gives a much lower feedback heating, allowing a large amount of cold gas with sufficient 
angular momentum to accumulate within the computational domain. 
The NFW+BCG simulation also uses $r_{\rm in}=0.5$ kpc but with a larger feedback efficiency ($\epsilon=5 \times 10^{-4}$),
and therefore the mass of the accumulated cold gas is reasonable. These numbers illustrate the weakness of the current feedback models in which 
the suitable $\epsilon$ depends, rather sensitively, on the radius at which $\dot{M}_{\rm in}$ is measured.

Instead of measuring $\dot{M}_{\rm in}$ at an arbitrary inner radius and tuning $\epsilon$ to get a reasonable 
match to observations as we are presently doing, our simulations suggest a more reliable way to estimate the SMBH accretion rate based on the mass distribution of low angular momentum 
cold gas (Eq. \ref{eq:Mdot_SCA} with $|\bm{l}| \lesssim 10^{28}$ cm$^2$s$^{-1}$). Table \ref{tab:tab1} shows that $\dot{M}_{\rm SCA}$ for
the NFW run with $r_{\rm in}=0.5$ kpc is about 6.4 times larger than the NFW run with $r_{\rm in}=1$ kpc but $\dot{M}_{\rm in}$ is about a factor of two
smaller for the former. This implies that $\dot{M}_{\rm BH}$ estimate based on the angular momentum distribution of cold gas is more robust because 
a larger mass accretion rate (as obtained from SCA) will prevent excessive cooling and mass deposition seen in 
the NFW run with $r_{\rm in}=0.5$ kpc.
Therefore, we anticipate similar outcomes (similar cold gas mass, small scale accretion rate, etc.) for feedback simulations with identical 
$\epsilon$s if the SMBH accretion rate is estimated using the angular momentum distribution of the cold gas. In future we plan to carry out such 
simulations.

While the results in this section strongly suggest the importance of stochastic cold feedback in maintaining thermal equilibrium in cluster cores, our simulations have
 several limitations like the absence of star formation, magnetic fields and anisotropic thermal conduction. In the absence of stellar feedback and gas consumption 
 by star formation we see a build up of a massive cold gas torus. The next section explores the implications of cold gas depletion due to star formation,  
 via post-processing using a simple model.    In the following section, we also compare our 
 cold mode feedback results with Bondi/hot model and with observations.

\section{Cold versus hot feedback: simulations confront observations}
\label{sec:obs}

In this section we compare the results of cold feedback simulations with salient observations.

\subsection{Gas depletion due to star formation}

\begin{figure}
        \includegraphics[scale=0.38]{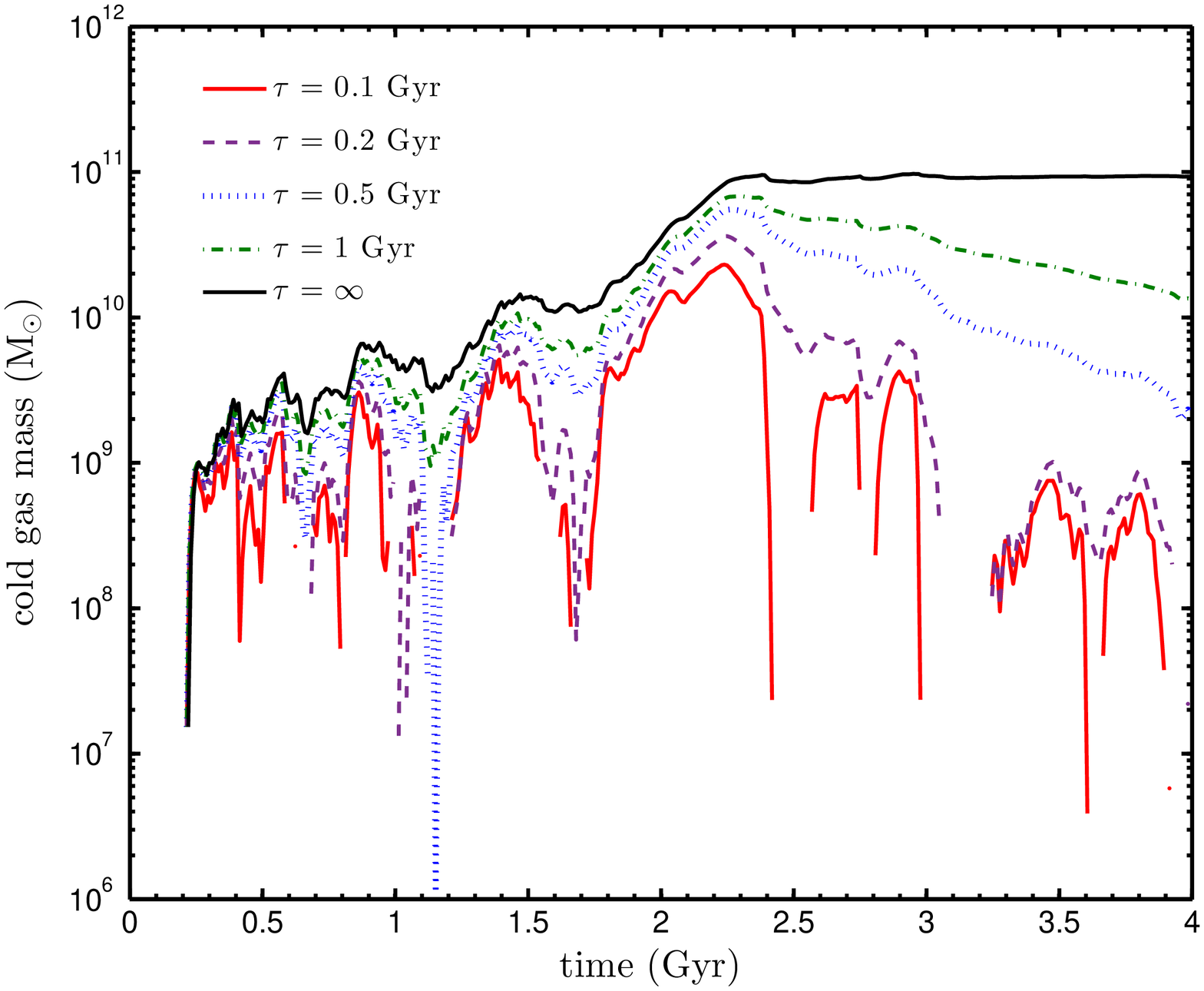}
        \caption{Cold gas mass as a function of time in the fiducial NFW run for different mass depletion time scales due to star formation.
        Without depletion of cold gas due to star formation ($\tau=\infty$; this is the same as the cold gas mass as a function of time in 
        Fig. 9 of \citetalias{pra15}), 
        cold gas secularly builds up with time. Depletion of cold gas due to star
        formation leads to cycles in cold gas mass as a function of time.
        }
        \label{fig:mcold_star}
\end{figure}

 Unlike \citet{li15}, our numerical simulations do not allow for the depletion of cold gas due to star formation. 
 This leads to a secular build up of cold gas to unrealistically large values. We can account for the influence 
 of star formation on cold gas depletion using a simple post-processing
model. The cold gas mass in presence of cold gas depletion due to star formation is given by
\begin{equation}
\label{eq:cold_mass}
\frac{d M_{\rm cold}}{dt} = \dot{M}_{\rm cond} - \frac{M_{\rm cold}}{\tau},
\end{equation}
where $M_{\rm cold}$ is the total cold mass in the computational domain, $\dot{M}_{\rm cond}$ is the rate of condensation of cold gas from the hot ICM 
(i.e., the derivative of the $\tau=\infty$ line in Fig. \ref{fig:mcold_star}), and $\tau$ is the cold gas depletion time scale (due to star formation). 
Therefore, the star formation rate
$dM_\star/dt = M_{\rm cold}/\tau$. While in reality, star formation and associated feedback will impact the surrounding gas distribution, we do not take this into account
as stellar feedback is subdominant on global scales in cluster cores (although it may help drive local turbulence in the cold gas; e.g., \citealt{hob11}). For a constant $\tau$, Eq.
\ref{eq:cold_mass} can be solved analytically to give
\begin{equation}
M_{\rm cold}(t) = \int_0^t e^{-(t-t^\prime)/\tau} \dot{M}_{\rm cond}(t^\prime) dt^\prime.
\end{equation}
Figure \ref{fig:mcold_star} shows the cold gas mass as a function of time for different cold gas mass depletion time scales ($\tau$). For a gas depletion 
time scale $\lesssim 0.5$ Gyr the cold gas mass is always $\lesssim 5\times10^{10} M_\odot$ and most of the gas is channeled into stars. For $\tau \lesssim 0.2$ Gyr, there are times when there is no cold gas present in the cluster core. Cold gas mass in our simulations is on the higher side and would lead to an average star 
formation rate of $\approx 10^{11} M_\odot/({\rm 5~Gyr}) = 20$ M$_\odot$yr$^{-1}$. This rate is comparable to that seen in some of the BCGs in cool core clusters (e.g., \citealt{bil08,mit15,lou16}) but is on the higher side of the the observed distribution.
This is understandable because in \citetalias{pra15} we deliberately 
chose the accretion efficiency of
$\epsilon=6\times10^{-5}$, the minimum value required to bring down the cooling rate by a factor of 0.1 with respect to cooling flow in absence of AGN feedback. This suggests that the feedback efficiency 
(with respect to the accretion rate at 1 kpc) in typical clusters is between $10^{-4}-10^{-3}$. Accretion efficiency much above $10^{-3}$ (as in \citealt{gas12,li14}) leads to strong feedback that wipes out cold gas at 
radii $>10$ kpc at late times, and maintains the cluster in the hot state for most of the time. 

\begin{figure}
        \includegraphics[scale=0.37]{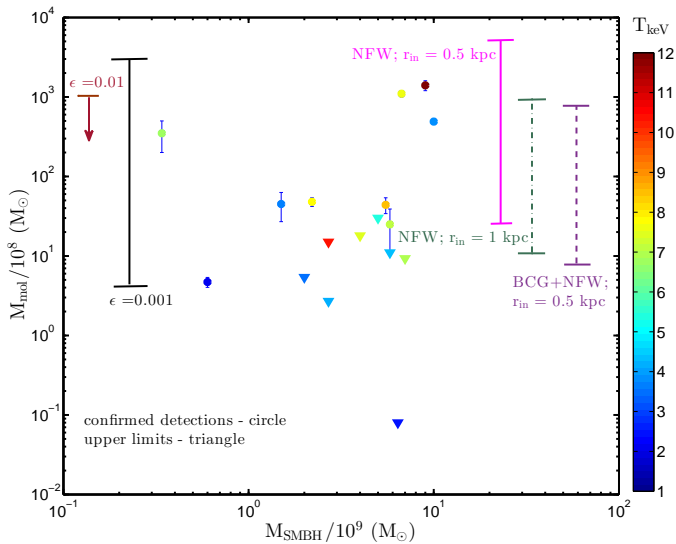}
        \caption{Molecular gas mass in centres of cool groups and clusters, obtained using CO observations, as a function of the SMBH mass.
        The filled circles represent detections and the triangles are upper limits.
        The cluster temperature is represented by color (obtained from \citealt{cav09}). There is a large scatter in the molecular gas mass for a
        given SMBH mass. Three vertical lines on extreme right denote the range of cold gas mass in our three simulations (measured from 1 Gyr till the end of the run,
        without accounting for cold gas depletion). A vertical line
        and an upper limit in extreme left are obtained from the simulations of \citet{li15} with their feedback efficiency of 0.01 and 0.001
        (bottom panels of their Figs. 2 \& 7).  
        Our model for cold gas depletion (see Eq. \ref{eq:cold_mass}  \& Fig. \ref{fig:mcold_star}) can also reduce the cold gas mass (of course at
        the expense of creating more stars).
        }
        \label{fig:mcold_vs_MBH}
\end{figure}

\subsection{Cold gas mass, SMBH and halo mass}
\label{sec:SMBH_halomass}
The mass accretion rate onto the SMBH in hot/Bondi accretion is strongly dependent on the mass of the SMBH (Eq. \ref{eq:MdotB}).
In the standard Bondi scenario, one does not normally consider the gas to cool to very low temperatures (e.g., \citealt{qua00}). Cold gas in
cluster cores can be naturally understood as a result of local thermal instability in the core satisfying 
rough global thermal balance. Under the assumption that the cold gas mass in the Bondi scenario is proportional to the SMBH accretion rate, 
the cold gas mass in cluster cores
is expected to be $\propto M_{\rm BH}^2 n T_{\rm keV}^{-3/2}$ (i.e., strongly dependent on SMBH and halo properties).
Figure \ref{fig:mcold_vs_MBH}, adapted from \citet{mcn11}, shows the cold gas mass as a function of the SMBH mass for different groups and clusters. The scatter in cold gas mass is very large given the rather uniform core entropies observed in cool cores, unlike the expectation from the Bondi accretion rate. Also the observations do not show a strong correlation (anticorrelation) between the SMBH mass 
(cluster temperature; indicated by the colors of markers) and the mass of molecular gas.
In fact, the largest molecular gas mass occurs in the hottest cluster, unlike what is expected from $T_{\rm keV}^{-3/2}$ 
scaling of $\dot{M}_B$ (Eq. \ref{eq:MdotB}; this assumes that the
temperature of the hot gas at the Bondi radius scales with the core X-ray temperature; X-ray spectra do not show X-ray emitting gas below 0.3 the
cluster temperature; e.g., \citealt{pet03}; see also Fig. 1a of \citealt{hog16}). Similarly, in Figure \ref{fig:mcold_vs_MBH} the data point with 
the smallest SMBH mass has a large cold gas mass.
There is a large scatter in cold gas mass, irrespective of the SMBH mass and the cluster temperature, a hallmark of cold mode feedback
that is stochastic and shows cooling and heating cycles (\citealt{pra15,li15}).

Figure 9 in \citetalias{pra15} and Figures 2 \& 7 in \citet{li15} show that the cold gas mass, for the same SMBH and halo, are highly time variable and show a broad distribution.  In fact, at times, one may not see any cold gas because star formation and stellar feedback can, in principle, consume all the cold gas before it is replenished by ICM cooling (this is seen in Fig. 2 of Li et al. 2015, who explicitly model star formation and stellar feedback in their simulations, as well as in our Fig. \ref{fig:mcold_star}, where we attempt to infer the impact of star formation, but not stellar feedback, on the cold gas mass in our simulation during post-processing).   The range of cold gas mass seen in our numerical simulations is indicated by the three vertical intervals on the right in Figure 8.   Even without star formation and stellar feedback, the lower end of the cold gas mass distribution in our simulations overlaps with the observed range of cold gas mass in cluster cores.   Typically, however, the cold gas mass is too high by up to two orders of magnitude.   Allowing for gas depletion by star formation on timescales of $0.2$--$0.5$ Gyr during post-processing (Fig. \ref{fig:mcold_star}) brings our gas mass into agreement with the observations.  Nonetheless, this is an issue that requires further study, and we are in the process of explicitly including a realistic treatment of star formation and stellar feedback
, as well as allowing for the effects of quasar-mode feedback that may occur when the AGN occasionally switches to that mode, 
in our simulations.

\begin{figure}
        \includegraphics[scale=0.37]{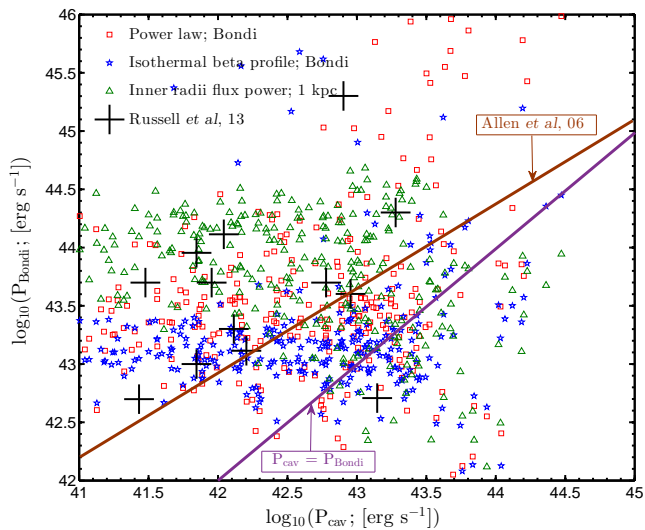}
        \caption{ Bondi power (estimated using two different extrapolations of the spherically angle averaged hot gas [$T>0.1$ keV] density profiles
        to the Bondi radius)
        as a function of cavity power from our fiducial NFW run. Simulation data is sampled every 10 Myr and there is a large
        scatter for both extrapolations. Also shown as green
        triangles is the instantaneous accretion power ($\epsilon \dot{M}_{\rm in} c^2$) supplied by the SMBH. At most times Bondi
        power is within $10^{43}-10^{45}$ erg s$^{-1}$, but cavity power has a much larger variation. A similar behaviour is seen
        in our other runs listed in Table \ref{tab:tab1}. Data points from \citet{rus13} and the best-fit from \citet{all06} are also shown. }
        \label{fig:Bondi}
\end{figure}

\subsection{Bondi power and cavity power}
\label{sec:Bondi}

\citet{all06} noted a tight correlation between the Bondi power and the cavity power for nearby elliptical galaxies and argued
for the validity of Bondi accretion. \citet{rus13} recently refined the determination of Bondi and cavities powers from Allen et al.'s sample
and included more data points, and found that the correlation is much weaker.
Figure \ref{fig:Bondi} shows the Bondi-power cavity-power relationship obtained for our fiducial NFW run where, as we have discussed, the jet is in fact triggered by cold gas condensation, not Bondi accretion. Bondi power
$P_B = 0.1 \dot{M}_B c^2$, where $\dot{M}_B$ is given by Eq. \ref{eq:MdotB}, is calculated using the 
hot gas density profile in our domain, extrapolated to the Bondi radius, 
\begin{equation}
\label{eq:RB}
R_B \equiv 2 GM_{\rm BH}/c_{s, \infty}^2 \approx 35~{\rm pc}~ M_{\rm BH, 9} T_{\rm \infty,  keV}^{-1}.
\end{equation}
 Emulating what \citealt{rus13} do with the observational data, we also use two different best-fit models for the hot gas density applied from $r_{\rm in}$ to 10 kpc:
a power-law [PL] model (red squares in Fig. \ref{fig:Bondi}); and a $\beta$ model (blue stars) with the hot gas
density going like $n_0 [1+(r/r_0)^2]^{-3\beta/2}$.
A power-law profile gives a larger Bondi accretion rate (see Eq. \ref{eq:MdotB}) because
the density extrapolated to the Bondi radius is higher.
The temperature at the Bondi radius (needed to calculate $c_{s, \infty}$ in Eq. \ref{eq:RB}) is assumed
to be the same as that at $r_{\rm in}$; results are unchanged with other reasonable extrapolations of the temperature.
 Also shown in Figure \ref{fig:Bondi} is the instantaneous value of the feedback power, where
$\epsilon \dot{M}_{\rm in} c^2$ (green triangles), $\dot{M}_{\rm in}$ is the total [hot+cold] accretion rate measured at the inner radius $r_{\rm in}$.

{\it Estimating cavity power:} The cavity power in Figure \ref{fig:Bondi} is calculated in the usual way. To estimate the jet/cavity energy we compute $4\int pdV $ over the volume of 
low density regions (density lower than 0.2 times the initial minimum density; results are insensitive to the exact value of the threshold as 
long as it is smaller than the initial minimum density), corresponding to the enthalpy of a relativistic bubble. To convert this into power, 
we divide the energy by an estimate of the jet sound crossing time using the outermost radius of the bubble material
($r/[600~{\rm km~s}^{-1}]$; 600 km s$^{-1}$ is the sound speed for the 1.5 keV cluster core plasma; see the bottom-right panel 
in Fig. 2 of \citetalias{pra15}). This definition is slightly different from what was used in \citetalias{pra15}, but the results are
independent of the details of jet power and Bondi estimates.  As discussed in \citetalias{pra15}, there is a large scatter 
between cavity power and the instantaneous accretion power because 
of hysteresis in cold mode feedback. 

More importantly, the scatter and the lack of obvious correlation in the observational data from \citet{rus13} in Figure \ref{fig:Bondi}
is similar to that for the Bondi power-cavity power scatter seen in our fiducial simulation. Notice that the variation in Bondi power is smaller 
than the variation in cavity power in the simulations.
This is because the hot gas properties in the core (on which the Bondi estimate depends) do not vary as much in time as the cold gas 
accretion rate. Moreover, cooling and accretion events occur closely spaced in time and our feedback prescription can blow really 
powerful cavities (see jet power in Fig. \ref{fig:mdot_vs_time}).
The observational data point in Figure \ref{fig:Bondi} with the highest Bondi power is for M87, for which \citet{rus13} use
a SMBH mass of $6.6 \times 10^{9} M_\odot$, much larger than our fiducial choice of $10^9 M_\odot$; recalling that
$\dot{M}_B \propto M_{\rm BH}^2$ brings this point in agreement with the simulation results. 
The highest cavity power scales with the halo mass and our halo mass is on the higher side. 
Also, unlike simulations, observations can miss out fainter parts of cavities, leading to an underestimate of the cavity power. 

\section{Discussion}
\label{sec:astro}

The problems with Bondi accretion (hot accretion in general) are well recognised (see section \ref{sec:intro}). Standard viscous mediated 
accretion flows onto SMBHs, even with a large $H/R\sim 0.1$, leads to the angular momentum problem; namely, the viscous accretion time 
scale for the accretion flow is much longer than the cooling time of the cluster core. Therefore, enough gas cannot accrete fast enough 
and provide sufficient feedback power to prevent a strong cooling flow.  

A number of published studies have invoked the stochastic nature of the condensing cold gas and its fundamentally different behaviour 
compared to the standard accretion flow 
 to overcome the angular momentum problem (\citealt{piz10,hob11,nay12,gas13}). Through 3-D hydrodynamic simulations, we {\sl explicitly} demonstrate that 
 the angular momentum of cold gas condensing out of the hot phase indeed varies stochastically and has an isotropic distribution 
 (especially for the low specific angular momentum, $l \lesssim 10^{28}$ cm$^2$s$^{-1}$, cold gas). Cancellation of almost randomly directed 
angular momentum on short timescales leads to the formation of a compact accretion flow that accretes onto the SMBH on the much 
shorter dynamical time scale (compare Eqs. \ref{eq:Mdot} \& \ref{eq:Mdot_SCA}). 
Recent ALMA observations of infalling cold, clumpy CO clouds in the elliptical galaxy at the centre of Abell 2597 (\citealt{tre16}), within  
$100$ pc of the SMBH,  provides an almost direct evidence for stochastic cold accretion (SCA). 
\citet{nay12} go onto argue that even in the higher $\dot{M}$ regime, SCA is responsible for the emergence of the SMBH $M_{\rm BH}-\sigma$ 
relation because accretion via a standard 
extended disk results in more massive SMBHs than observed since a massive extended thin disk does not couple efficiently to an isotropic feedback-powered outflow. Turbulence due to stochastic angular 
momentum and infall of cold gas (viscous time is much shorter for a turbulent flow with $H/R \sim 0.1$) also prevents the problem of gravitational 
fragmentation in stochastic accretion flow.

Most of the earlier studies of stochastic cold accretion were carried out in numerical setups with turbulence in the cold gas driven 
artificially (to realise the stochastic distribution of cold gas angular momentum). However, in this paper we have shown that the turbulence emerges self-consistently from jet-ICM interactions and this behaviour is sustained over timescales of several Gyr (i.e., much longer than the cluster core cooling time). We analyze the angular momentum distribution of cold gas 
in our AGN jet-ICM simulations (which appear consistent with most observations of cool cluster cores; see section \ref{sec:obs}) and demonstrate 
that stochastic cold accretion is indeed realised in cool cluster cores stirred by time-dependent AGN jets.

In cluster simulations with limited resolution,
the accretion physics input necessarily has to be inserted via a sub-grid 
model. One such model, which is much better than what is typically used (including ours in this paper where we estimate $\dot{M}$ at $\sim 1$ 
kpc and multiply it by an efficiency parameter $\epsilon$ to calculate the mechanical power injected by the SMBH), is to use the mass accretion 
rate of the low angular momentum gas (with say $l \lesssim 10^{28}$ cm$^2$s$^{-1}$; see Eq. \ref{eq:Mdot_SCA} \& Fig. \ref{fig:mdot_vs_time}) 
crossing  
the smallest resolved radius (see section \ref{sec:improv} for details). 
Another improvement is the inclusion of star formation and realistic stellar feedback (stellar feedback is increasingly important for lower mass
halos such as clusters and elliptical galaxies; e.g., see Fig. 4 in \citealt{sha12b})  
that can prevent artificial accumulation of massive
cold gas in cluster cores (e.g., as done by \citealt{li15}). Also, as mentioned earlier, our choice of $\epsilon$ (feedback efficiency parameter) is
on the lower side, resulting in somewhat large mass deposition and star formation in the core. More 3-D simulations with larger $\epsilon$s
are required to get a complete picture.

\subsection{Uncertainties in accretion physics}

Much more needs to be done in understanding the details of stochastic cold accretion (SCA). In our simulations, which focus on larger 
($\gtrsim 1$ kpc)
scales, we have to make assumptions about the underlying stochastic accretion. In particular, we need to understand how the stochastic 
cold gas condensing out of the ICM is eventually accreted onto the SMBH. Whether a disk forms
after the cancellation of stochastic angular momentum and on what timescale, and at what radii does the transition from SCA to a viscous 
thin disk (if it does at all) and to 
a RIAF (radiatively inefficient accretion flow) occurs. While the transition radius from a thin disk to an inner hot/thick RIAF occurs at a
smaller radius for a higher accretion rate, the details are  not clear even for the 
standard (non-stochastic) accretion flows with a fixed angular momentum axis (e.g., see the vastly different estimates of disk to RIAF transition 
radius in \citealt{das13} [Fig. 9b], \citealt{liu99} [Fig. 1], \citealt{yua14} [Fig. 7b]). 

Strong AGN jets and X-ray cavities (which are powered by jets) are expected only in the radiatively inefficient accretion 
regime for which $\dot{M}_{\rm BH} \lesssim \alpha^2 \dot{M}_{\rm Edd}$ (\citealt{ree82,chu05,nem07}). Our SMBH accretion rate at most times 
is lower than this threshold for jet formation (see Fig. \ref{fig:mdot_vs_time}). Infrequent, short duration high $\dot{M}_{\rm BH}$ events, in fact, 
provide a potential basis for short-lived transition from radio mode to quasar mode and for changing the direction of the jet, as observed in numerous clusters (e.g., \citealt{bab13}). 
Generally the mass accretion
rate onto the SMBH is expected to be even lower because of outflows in RIAFs (larger suppression is expected for bigger RIAFs).
Therefore, advances in accretion physics are required
to estimate the size of RIAFs and the mass accretion rate onto SMBHs fed by SCA. 

To sum up, majority of the cold gas condensing out of the ICM due to thermal instability has large angular momentum such that the viscous 
accretion timescale is longer than the core cooling time for $l \gtrsim 10^{28}$ cm$^2$s$^{-1}$ (Eq. \ref{eq:tvisc}). Most of this gas is
expected to be Toomre
unstable and to form stars, instead of 
being accreted onto the SMBH. The lower angular momentum cold gas ($l \lesssim 10^{28}$ cm$^2$s$^{-1}$) has stochastic angular 
momentum, which implies that cloud collisions (which lead to rapid angular momentum cancellation) will cause the
cold, turbulent gas to be accreted rapidly, without much gravitational fragmentation. At innermost radii typically a RIAF with powerful jets is expected, 
since $\dot{M}_{\rm BH} \ll 0.01 \dot{M}_{\rm Edd}$. 

\section{Summary}
\label{sec:summary}
We finally summarise our major findings in the following points:
\begin{itemize}
\item Stochastic cold accretion (SCA), which was anticipated to play a key role in accretion onto SMBHs from idealised simulations, 
is realised naturally
in our realistic AGN jet-ICM simulations due to the turbulence induced by jets in a non-uniform ICM  over several Gyr time scale.

\item We find that the low angular momentum cold gas ($|\bm{l}| \lesssim 10^{28}$ cm$^2$s$^{-1}$) condensing out of the ICM 
has an isotropic distribution of angular momentum, which we expect will result in the cancellation of angular momentum on almost a dynamical time. 
This is unlike a  viscosity-mediated standard accretion flow, in which an extended thin disk with a long accretion time is expected to form.
Another advantage of a stochastic accretion flow is that the net angular momentum of the low angular momentum cold gas varies on a timescale shorter than the core
cooling time, implying that angular momentum cancellation and feedback driven by cold gas accretion can respond fast enough. Furthermore, 
SCA is independent of disk parameters like $\alpha$ and $H/R$. Elimination of these parameter (usually with a large range) makes SCA a simpler and more 
robust model.

\item Our work suggests an improved feedback prescription based on the angular momentum distribution of cold gas (section \ref{sec:improv}), which may be more robust than
the usual models in which the mass accretion rate is estimated at $\sim$ kpc scales. In the latter approach one has to fine tune the feedback efficiency parameter 
($\epsilon$) for different radii at which the mass accretion rate is estimated
(compare our NFW runs with the inner radius of 1 and 0.5 kpc).

\item Most of the features of cold accretion simulations, which show multiphase cooling and heating cycles driven by accretion of cold gas, 
match observations such as a large scatter 
of cold gas mass when compared to the SMBH and halo masses (this is not expected for hot/Bondi accretion; see section \ref{sec:SMBH_halomass}).   A large scatter observed between Bondi power and jet/cavity power is also consistent with our cold feedback simulations (see section \ref{sec:Bondi}).

\end{itemize}

\section*{Acknowledgments}
This work is partly supported by the DST-India grant no. Sr/S2/HEP-048/2012 and an India-Israel joint research grant (6-10/2014[IC]). 
DP is supported by a CSIR grant (09/079[2599]/2013-EMR-I). AB acknowledges support from NSERC, through the Discovery Grant Program, 
the Institut Lagrage de Paris, and Pauli centre for Theoretical Studies ETH UZH.  He also thanks the Institut d'Astrophysique de Paris (IAP), at the Institute for Computational Sciences and University of Zurich for 
hosting him. We acknowledge the support of the Supercomputing Education and Research Centre (SERC) at IISc for facilitating
our use of Cray XC40-SahasraT cluster on which some of our runs were carried out. We thank the anonymous referee for critical comments that
helped improve our paper. PS thanks Sagar Chakraborty for useful discussions.

\bibliographystyle{mnras}

\end{document}